\documentclass[
 reprint,
 amsmath,
 amssymb,
 aps,
 superscriptaddress,
 longbibliography,
]{revtex4-2}
\usepackage{CJKutf8}

\usepackage[pdftex]{graphicx}
\usepackage{wasysym}
\usepackage{amsfonts}
\usepackage{ifsym}
\usepackage{graphicx}
\usepackage{dcolumn}
\usepackage{bm}
\usepackage{hyperref}
\usepackage[normalem]{ulem}
\hypersetup{
    colorlinks=true,       
    linkcolor=blue,          
    citecolor=magenta,        
    filecolor=red,      
    urlcolor=cyan           
}
\usepackage{mathtools}
\usepackage{caption}
\usepackage{subcaption}

\usepackage{xcolor}
\definecolor{mBlue}{RGB}{31,119,180}
\definecolor{mOrange}{RGB}{255,127,14}
\definecolor{mGreen}{RGB}{44,160,44}
\definecolor{mRed}{RGB}{214,39,40}
\definecolor{mPurple}{RGB}{148,103,189}
\definecolor{mPink}{RGB}{227,119,194}
\definecolor{mBrown}{RGB}{140,86,75}

\newcommand{\COMMENT}[1]{}

\newcommand{\neqa}{\nonumber\end{eqnarray}}

\newcommand{\<}{{\langle}}
\renewcommand{\>}{{\rangle}}

\newcommand{\re}{\relax{\rm I\kern-.18em R}}

\def\su2{{SU(2)}}

\def\[{\left[}
\def\]{\right]}

\def\({\left(}
\def\){\right)}
\def\[{\left[}
\def\]{\right]}

\def\<{\langle}
\def\>{\rangle}

\def\i2{\frac{i}{2}}

\def\2F1{\,_2{\rm F}_1}

\newcommand{\beq}{\begin{equation}}
\newcommand{\eeq}{\end{equation}}
\newcommand{\beqq}{\begin{equation*}}
\newcommand{\eeqq}{\end{equation*}}
\newcommand\beqa{\begin{eqnarray}}
\newcommand\eeqa{\end{eqnarray}}
\newcommand\beqaa{\begin{eqnarray*}}
\newcommand\eeqaa{\end{eqnarray*}}
\newcommand\bea{\begin{array}}
\newcommand\eea{\end{array}}

\begin{document}
\begin{CJK*}{UTF8}{gbsn}

\title{Complexity Measure Diagnostics of Ergodic to Many-Body Localization Transition}

\author{Khen Cohen}
\affiliation{School of Physics and Astronomy, Tel Aviv University, Ramat Aviv 69978, Israel}
\author{Yaron Oz}
\affiliation{School of Physics and Astronomy, Tel Aviv University, Ramat Aviv 69978, Israel}
\author{De-liang Zhong (钟德亮)}
\affiliation{Blackett Laboratory, Imperial College London SW7 2AZ, United Kingdom}

\begin{abstract}

We introduce new diagnostics of the transition between the ergodic and many-body localization phases, which are based on complexity measures defined via the probability distribution function of the Lanczos coefficients of the tri-diagonalized Hamiltonian.
We use these complexity measures to analyze the 
power-law random banded matrix model as a function of the correlation strength and show that the moments and the
entropy of the distribution diagnose the ergodic to many-body transition, as well as the distinctive feature 
of the phases concerning the memory of the initial conditions.

\end{abstract}
\maketitle
\end{CJK*}

\section{Introduction}

The classification of phases of matter along with the identification of their diagnostics
is a fundamental scientific challenge.
Generically, isolated many-body quantum systems furnish two phases, the ergodic phase that exhibits late-time thermalization and the many-body 
localization (MBL) phase \cite{PhysRevLett.95.206603,Basko_2006} that fails to reach thermal equilibrium and retains a memory of the initial conditions (for a review see \cite{Nandkishore_2015,Abanin_2019,Sierant:2024khi}).
One diagnostic that distinguishes between the two phases is the entanglement entropy quantum information measure calculated by partitioning the system into two subsystems $A$ and $B$ and tracing out the degrees of freedom of one of them, say $B$.
The ergodic phase exhibits a large entanglement spreading and the entanglement entropy scales as the volume
of the subsystem $A$ (volume law), while the MBL phase exhibits less spreading of the entanglement, and the entanglement entropy scales as the area of the boundary of $A$ (area law).
Another diagnostic is a measure of the chaotic structure of the systems
quantified by the level spacing eigenvalue statistics of the Hamiltonian, or alternatively the $r$-statistics \cite{avishai2002level,Oganesyan_2007,oganesyan2009energy,pal2010many,iyer2013many,luitz2015many,regnault2016floquet,geraedts2016many,bertrand2016anomalous}, that
follow a random matrix ensemble distribution in the ergodic phase and a Poisson 
ensemble distribution in the MBL phase.

A complexity analysis approach to study the dynamics of quantum systems 
\cite{Parker:2018yvk, Barbon:2019wsy, Rabinovici:2020ryf, Balasubramanian:2022tpr} is rooted in the recursion method for solving the Schr\"{o}dinger equation \cite{Recursion_Method}
and the tri-diagonalization of the Hamiltonian of the physical system rather than its diagonalization. This has been successfully applied to diverse physical systems, including quantum systems with finitely many degrees of freedom, quantum field theories, and holographic theories \cite{Jian:2020qpp, Rabinovici:2021qqt, Bhattacharjee:2022vlt, Hornedal:2022pkc, Trigueros:2021rwj, Noh_2021, PhysRevA.105.L010201, Caputa:2022eye, Kar:2021nbm,  PhysRevA.105.062210, PhysRevResearch.4.013041, Bhattacharya:2022gbz, PhysRevD.104.L081702, Caputa:2021ori, Caputa:2021sib,Balasubramanian:2022tpr, Muck:2022xfc, Adhikari:2022oxr, Adhikari:2022whf,Rabinovici:2022beu,Rabinovici:2023yex,Rabinovici:2021qqt,Caputa:2023vyr,Caputa:2024vrn,Balasubramanian:2023kwd, Balasubramanian:2022dnj,Abanin_2019, Dymarsky:2019elm, Avdoshkin:2019trj,Menzler:2024atb,Camargo:2023eev,Huh:2023jxt,Dymarsky:2021bjq}.

Our aim in this Letter is to introduce new diagnostics of the two phases and the transition between them, which are based on complexity measures
defined via the probability distribution function of the Lanczos coefficients
that arise in the tri-diagonalization of the Hamiltonian.
We will argue that the moments and the
entropies of the distribution diagnose the ergodic/MBL transition and its distinctive feature 
concerning the \textit{memory of the initial conditions}. 
We will study the 
power-law random banded matrix (PRBM) model \cite{Mirlin_1996}, calculate the complexity
measures and showcase the ergodic/MBL transition. 

The existence of different diagnostics of the ergodic to MBL transition indicates that
there are several different properties that distinguish the two phases, 
corresponding to quantum information, quantum chaos, and the complexity of the dynamics.

The Letter is organized as follows.
In section II we provide the framework for the analysis including a brief discussion of the Krylov basis, the 
Lanczos coefficients and the PRBM model. In section III we introduce complexity measures based 
on the distribution of the normalized Lanczos coefficients and apply them to the analysis of the ergodic/MBL transition in the PRBM model.
Section IV is devoted to conclusions and outlook.
Details and additional numerical results are given in the supplemental materials.

\section{Framework}

\subsection{Krylov Basis and Lanczos Coefficients}

Consider a time-independent quantum Hamiltonian $H$ realized as a Hermitian matrix in a Hilbert space ${\cal H}$.
Starting from an initial state $|\psi_0\rangle \in {\cal H}$ one constructs a sequence of states by repeated actions of the Hamiltonian,
\beq \label{eqn-IterationBasis}
|\psi_1 \rangle = H |\psi_0\rangle, \quad |\psi_2 \rangle = H^2 |\psi_0\rangle, \ \cdots \ .
\eeq
The states $|\psi_\ell \rangle$ encode the evolution in time $|\psi(t)\rangle = e^{itH}|\psi_0\rangle$.
When $H$ is finite-dimensional one keeps acting with $H$ until we get a state that is a linear combination of the previous ones, and for a generic initial state, 
this process will terminate at the order of the dimension of ${\cal H}$.

The states $|\psi_\ell \rangle$ are neither normalized nor orthogonal. 
One performs a Gram-Schmidt procedure and gets an orthonormal set of states called \textit{Krylov} basis, denoted by $|\mathcal{K}_n\rangle$. The first state in the basis coincides with the initial state, $|\mathcal{K}_0 \rangle = |\psi_0\rangle$. The Krylov basis is a basis of a subspace of  ${\cal H}$ that depends
both on $|\psi_0\rangle$ and on $H$.

The Hamiltonian on this basis has a tri-diagonal form:
\begin{equation} \label{eqn-Ham-Kbasis}
H=\left(\begin{array}{cccccc}
a_1 & b_1 & 0 & 0 & 0 & 0 \\
b_1 & a_2 & b_2 & 0 & 0 & 0 \\
0 & b_2 & \cdots & \cdots & \ldots & \ldots \\
\ldots & \cdots & \cdots & \ldots & \ldots & b_{N-1} \\
\cdots & \cdots & \cdots & \ldots & b_{N-1} & a_N
\end{array}\right) \ ,
\end{equation}
and the coefficients $a_\ell$ and $b_\ell$ are called Lanczos coefficients. The diagonal elements of the Hamiltonian
$a_\ell$  vanish
when it is invariant under the $Z_2$ transformation $H\rightarrow -H$. In practice, all $b_\ell$ coefficients can be taken to be positive by a redefinition of the basis $|\mathcal{K}_\ell \rangle \rightarrow -|\mathcal{K}_\ell \rangle$.

The Lanczos coefficients have the dimension of energy and they carry information on the complexity of the physical systems and their evolution in time. For instance, 
it has been proposed \cite{Parker:2018yvk} that in the limit $\ell \gg 1 $, the Lanczos coefficients distinguish between chaotic and integrable systems: for the chaotic system, $b_\ell \sim \ell$, while 
for the integrable system, $b_\ell \sim 1$ \footnote{This proposal is imprecise and there exist counter-examples, see e.g. \cite{Bhattacharjee:2022vlt}.}. 

\subsection{Power-Law Random Banded Matrix Model} 

We will study the transition from the ergodic to the MBL phase in the 
power-law random banded matrix (PRBM) model employed in \cite{Mirlin_1996,PhysRevB.61.R11859,evers2000fluctuations,mirlin2000multifractality,evers2008anderson,kravtsov1997new,mendez2012multifractal,varga2002fluctuation,mendez2006scattering,mendezbermudez2014generalized,khemani2017two,rao2022power}. The Hamiltonian $H$ is an element of
the PRBM ensemble, represented by an $N \times N$ matrix whose $(ij)$ element is given by: 
\beq H_{ij} = \rho(|i-j|,\mu) \times a_{ij}\, . 
\label{H}
\eeq 
Here, each $a_{ij}$ is randomly sampled independently from a Gaussian distribution with a zero mean and a variance of one. 
The $\rho(r, \mu)$ is defined as follows: 
\begin{equation}
    \rho(r,\mu)  \coloneqq \begin{cases}
        1 & \text{if}\ r < \mathtt{B} \, , \\
        \big(1+r^{2\mu} \big)^{-1} & \text{if}\ r \geq \mathtt{B} \ ,
    \end{cases}
    \label{Cor}
\end{equation}
with $\mu \in \mathbb{R}_+$ being a free parameter that controls the strength of the correlation between
the Hamiltonian matrix entries. The bandwidth parameter $\mathtt{B} \in \mathbb{Z}_{>0}$ controls the interaction range of the model.

Analytics shows that in the large $\mathtt{B} \gg 1$ limit there is a phase transition at $\mu_c = 1$ \cite{Mirlin_1996,mirlin2000multifractality,evers2000fluctuations}. For small $\mathtt{B}$, there is no analytical result, but numerical observations indicate that for $\mathtt{B}=1$, this model undergoes a transition from ergodic to MBL phase in the vicinity of $\mu = 1$, as indicated by the r-statistics of the eigenvalues level distribution
and the entanglement entropy \cite{varga2000critical,carrera2021multifractal,rao2022power}. The special cases, $\mu=0$ and
$\mu \rightarrow \infty$, correspond to chaotic and integrable Hamiltonians, and
exhibit Gaussian Orthogonal Ensemble (GOE) and Poisson eigenvalues level spacing statistics, respectively.

In the following, we will introduce new complexity measure diagnostics of the ergodic/MBL transition based on the
Lanczos coefficients.

\section{Complexity Diagnsotics}

\subsection{Lanczos Coefficients Distribution (LCD)}

We define the averaged Lanczos coefficients distribution (LCD) as follows. We choose a fixed initial state $|\psi_0\rangle$ and sample Hamiltonian matrices from the 
PRBM distribution (\ref{H}),(\ref{Cor}). For each sampled Hamiltonian, we construct the Krylov basis as described in the previous section and read out the Lanczos coefficients $b^\mu_\ell$ from the Hamiltonian in that basis, see \eqref{eqn-Ham-Kbasis}. We repeat the sampling several times and evaluate the dimensionless normalized ensemble averages.
\footnote{Note, that the reason for not considering ensemble averages of the Lanczos coefficients $a_\ell$ is the reflection symmetry of the random matrix model $H \rightarrow -H$ makes them vanish. }
\begin{gather}\label{b} 
    \mathtt{b}^\mu_\ell \equiv \frac{\langle b^\mu_\ell \rangle}{\sum_{\ell=1}^{N-1} \langle b^\mu_\ell \rangle},~~~\ell=1,...,N-1 \ .
\end{gather}
According to the definition (\ref{b}), they implicitly depend on the initial state. Also, $b_\ell$ decreases with $\mu$ and there is no clear indication of the transition, if we do not normalize as in (\ref{b}); see appendix \ref{apd:Unnormalized} for the raw data.

When $\mu=0$ (GOE), there is no dependence on the initial state as all different initial states are related by $O(N)$ rotations
that do not affect the distribution. The analytical expression for the ensemble average of $b_\ell$ (\ref{b}) reads \cite{Dumitriu2002} 
\beq \label{ab}
\langle b_\ell^{\mu = 0} \rangle = \frac{\Gamma \left(\frac{N-\ell +1}{2}\right)}{\sqrt{2} \Gamma \left(\frac{N-\ell}{2}\right)} \ . 
\eeq 

The limit $\mu \rightarrow \infty$ corresponds to the integrable ensemble with a Poisson distribution of the Hamiltonian's eigenvalues
level spacing and vanishing Lanczos coefficients. 
However, the Lanczos coefficients averages (\ref{b}) for a finite number $M$ of sampled Hamiltonians (\ref{H})
in the integrable regime $\mu \geq \sqrt{M}$ exhibit a uniform LCD, as explained and verified numerically in appendix \ref{apd:Poisson}.

The distribution $\mathtt{b}^\mu_\ell$ has a compact support since $l\leq N$, which will be important for quantifying the
differences between its various moments as diagnostics of the transition. Note, that one can also take the continuum limit
of the distribution as the infinite limit of $l$ and $N$ keeping $x=\frac{l}{N}$ fixed
\cite{balasubramanian2022tale}, but we will work with the discrete LCD.

Based on the distribution $\mathtt{b}^\mu_\ell$ we can study different observables that capture the complexity of the system as a function of $\mu$, such as the (normalized) moments of the distribution:
\beq
\mathbb{E}^\mu_k \coloneqq \sum_{\ell=1}^{N-1} \left(\frac{\ell}{N} \right)^k \mathtt{b}^{\mu}_\ell\, ,
\eeq
and the R\'{e}nyi entropy,
\begin{equation} \label{eqn-RenyiE}
\mathbb{H}^\mu_\alpha \coloneqq \frac{1}{1-\alpha} \log \sum_{\ell=1}^{N-1} [\mathtt{b}^{\mu}_\ell]^\alpha\, ,
\end{equation}
which in the $\alpha \rightarrow 1$ limit is the Shannon entropy: 
\begin{equation}
\mathbb{H}^\mu \coloneqq -\sum_{\ell=1}^{N-1} \mathtt{b}^{\mu}_\ell \log \mathtt{b}^{\mu}_\ell \ .
\end{equation}

\subsection{PRBM Numerical Simulations }

\subsubsection{Initial States}

We have chosen seven typical normalized initial states to run the numerical simulations \footnote{The $\mu$ points that we use for the numerics are given by
\begin{multline*}
\mu \in \{0.0, 0.2, 0.4, 0.6, 0.8, 0.9, 0.95, 1.0, 1.05, 1.1, 1.15, \\ 1.2, 1.3, 1.4, 1.5, 1.6, 1.7, 1.8, 1.9, 2.0, 2.3, 2.5, 2.7, 3.0, 3.3, \\ 3.5, 4.0, 4.5, 5.0, 5.5, 6, 7, 8, 9, 10\}\, .
\end{multline*}
Unless specified otherwise, the data points colored in \textcolor{mBlue}{blue}, \textcolor{mOrange}{orange}, \textcolor{mPurple}{purple}, \textcolor{mGreen}{green}, 
    \textcolor{mPink}{pink},
    \textcolor{mBrown}{brown} and \textcolor{mRed}{red} correspond to the initial states $\textcolor{mBlue}{|\psi_1\rangle}, \cdots \textcolor{mRed}{|\psi_7\rangle}$, respectively, as defined in \eqref{eqn-IntialState}.
    The matrix size is $N=500$ and the ensemble consists of $M=9600$ samples.
}.
They include:
\beq \label{eqn-IntialState}
\begin{aligned}
\textcolor{mBlue}{|\psi_1 \rangle} & = (1,0,0,\cdots, 0)^\intercal, \\
\textcolor{mOrange}{|\psi_2 \rangle} & = \frac{1}{\sqrt{2}}(1,0,0,\cdots, 0, 1)^\intercal, \\
\textcolor{mPurple}{|\psi_3 \rangle} & = 
\frac{1}{\sqrt{N/4}} \underbrace{(1, 1, \cdots, 1, }_{N/4\ \text{times}} \underbrace{0, \cdots, 0)}_{3N/4\ \text{times}}{}^\intercal
 \\
\textcolor{mGreen}{|\psi_4 \rangle} & = 
\frac{1}{\sqrt{N/2}} \underbrace{(1, 1, \cdots, 1, }_{N/2\ \text{times}} \underbrace{0, \cdots, 0)}_{N/2\ \text{times}}{}^\intercal
 \\
 \textcolor{mPink}{|\psi_5 \rangle} & = 
\frac{1}{\sqrt{3N/5}} \underbrace{(1, 1, \cdots, 1, }_{3N/5\ \text{times}} \underbrace{0, \cdots, 0)}_{2N/5\ \text{times}}{}^\intercal
 \\
 \textcolor{mBrown}{|\psi_6 \rangle} & = 
\frac{1}{\sqrt{3N/4}} \underbrace{(1, 1, \cdots, 1, }_{3N/4\ \text{times}} \underbrace{0, \cdots, 0)}_{N/4\ \text{times}}{}^\intercal
 \\
\textcolor{mRed}{|\psi_7 \rangle} & = \frac{1}{\sqrt{N}} (1,1,\cdots, 1)^\intercal\, . \\
\end{aligned}
\eeq
These states have different structures of zero and one entries, which will be reflected in the LCD distribution 
of the MBL phase that keeps a memory of the initial state.

\subsubsection{LCD} \label{sec:LCD}

\begin{figure}[!h]
    \centering    \includegraphics[width=0.5\textwidth]{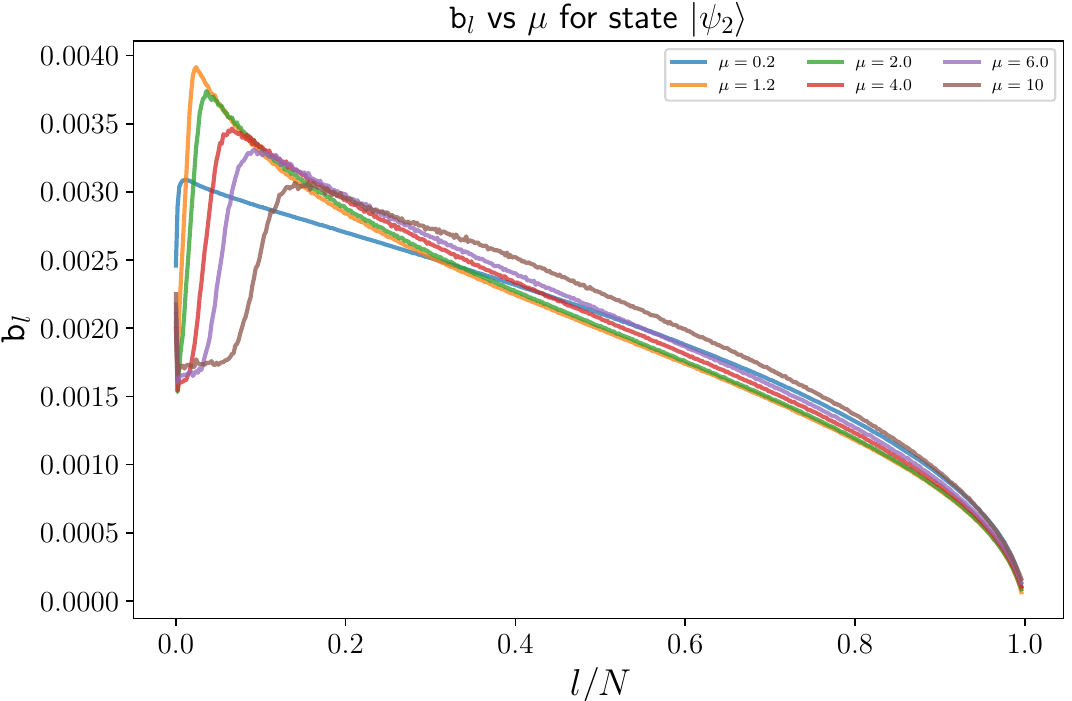}
    \caption{The LCD $\mathtt{b}^\mu_\ell$ corresponding to the state $|\psi_2\rangle$ for various values of $\mu$. Here the different colours represent different $\mu$ values rather than initial states. We see that the main difference is at low 
    $\frac{\ell}{N}$. A detailed error analysis is shown in appendix \ref{apd:ErrorAnalysis}.}
    \label{fig:LCD_psi2}
\end{figure}

For any finite-dimensional Hilbert space, the iteration process \eqref{eqn-IterationBasis} must terminate, so the LCD is a probability distribution with a compact support, $\ell/N \in [0,1]$.
In Fig. \ref{fig:LCD_psi2}, we plot the LCD for $\textcolor{mOrange}{|\psi_2\rangle}$ with six different values of $\mu$ (see also Figs. \ref{fig:LCD_psi_all} and \ref{fig:LCD_vs_mu} in appendix \ref{apd:LCD} for the LCD of the other initial states).
These values are selected to be before, match, and after the identified transition at $\mu_c \simeq 1.2$. We observe that the difference between the distributions at different $\mu$ comes mainly from the small $\ell/N$ range, while 
power-law tail region $\ell/N \sim 1$ is quite similar and state independent.
This structure of the LCD implies that while different moments can diagnose the transition,
the low moments which are more sensitive to the small $\ell/N$ region have a memory of the initial state 
feature of the MBL phase, see Figs. \ref{fig:LCD_psi_all} and \ref{fig:LCD_vs_mu} in appendix \ref{apd:LCD} for more details. In the literature, it is generally argued that the behavior of the large $\ell$ Lanczos 
coefficients $b_{\ell}$ quantifies the difference between integrable and chaotic systems. We see from the LCD 
distribution that in fact the small $\ell$ Lanczos coefficients
are more effective diagnostics of this.

\subsubsection{LCD Moments and Entropy}

\begin{figure}[hbt!]
    \centering    \includegraphics[width=\linewidth]{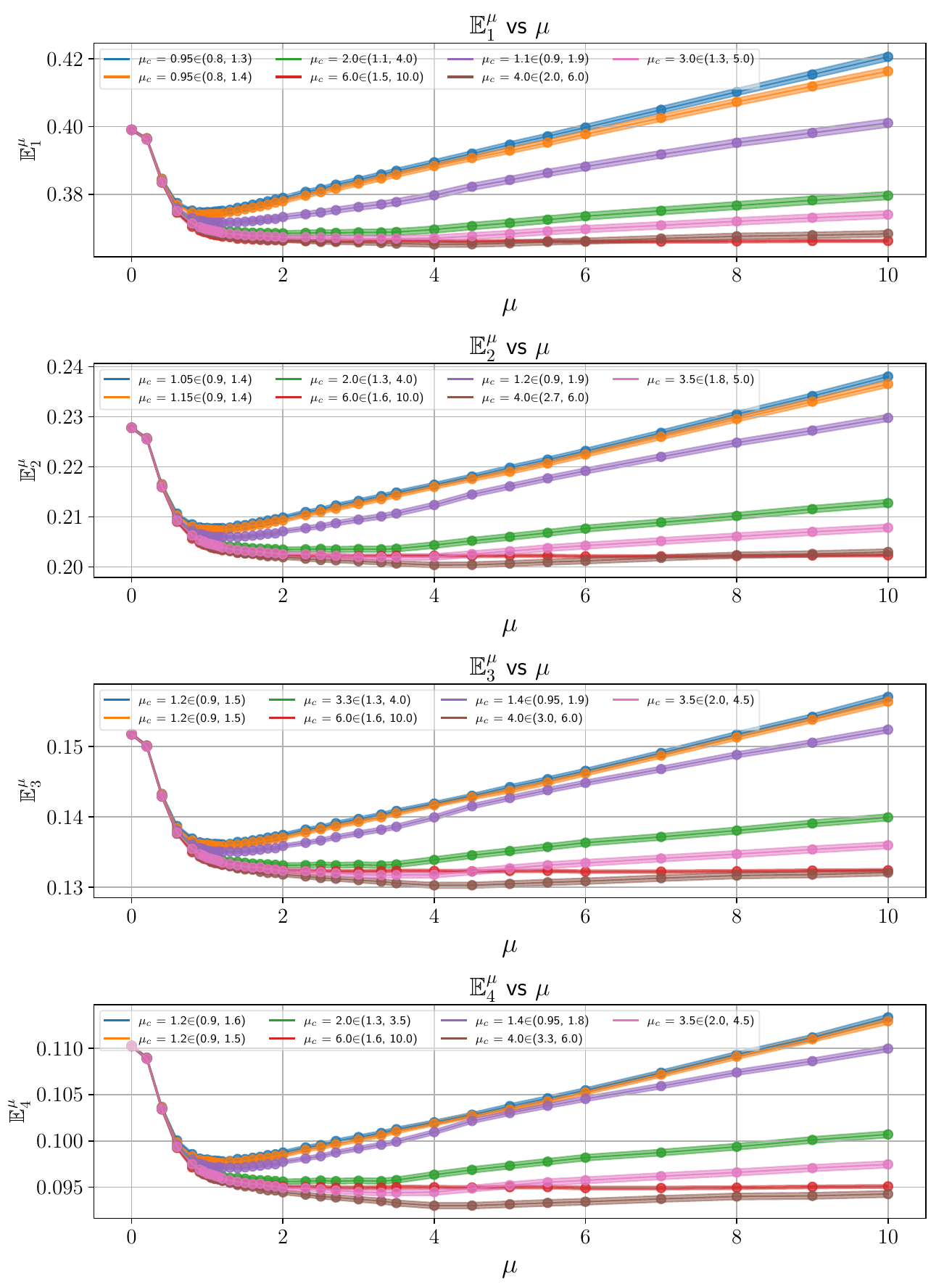}
    \caption{The first four LCD moments across different $\mu$ values and initial states with error bars. We see the transition from the ergodic to MBL phase, as well as the memory of the initial state at the MBL phase and the memory-less ergodic phase.}
    \label{fig:bn_states_moments}
\end{figure}

\begin{figure}[hbt!]
    \centering
    \includegraphics[width=\linewidth]{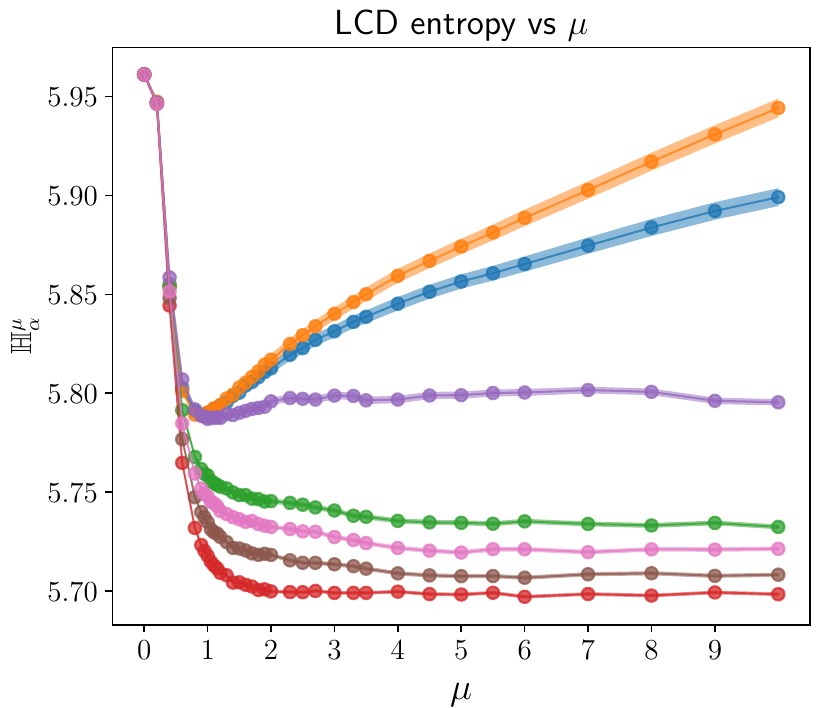}
    \caption{The LCD R\'{e}nyi entropy \eqref{eqn-RenyiE} evaluated at $\alpha =10$ across different $\mu$ values and initial states. We see the transition and the initial state dependence of the MBL phase. }
    \label{fig:bn_states_entropy}
\end{figure}

The first four moments of the LCD, along with the LCD entropy, are presented in Figs. \ref{fig:bn_states_moments} and \ref{fig:bn_states_entropy}, respectively. To enhance visual coherence, we have colored all initial states and 
ensured that the color of each initial state matched the color of the corresponding data points across all plots.

In both cases, the qualitative behavior exhibits notable similarities. For the case where $\mu < \mu_c \simeq 1.2$, the physical system exhibits a property of state independence: the LCD moments, as well as the LCD entropy, are identical for various initial states. The situation changes when $\mu > \mu_c$; here, the LCD moments and the LCD entropy begin to show a dependence on the initial states. This distinction can be understood in terms of the ergodicity of the physical system. Specifically, when the system is in the ergodic phase, which occurs before the transition at $\mu_c \simeq 1.2$, it remains unaffected by its initial state, as for the standard GOE ensemble case. However, after the transition, as $\mu$ exceeds $\mu_c$, the system enters a non-ergodic phase and the data starts to depend on the initial state. 
In this phase, the eigenstates' level spacing follows a semi-Poisson distribution. Consistent with the expectations for a gradual transition, the behavior of the moments post-transition varies depending on the chosen initial state.

Beyond the critical value $\mu_c \approx 1.2$, in the non-ergodic regime, the system undergoes a non-sharp phase transition to the Poisson distribution and transitions into an intermediate phase, as found in \cite{PhysRevB.93.041424, PhysRevB.99.104205}. In this phase, the level-spacing statistics of the system follow a semi-Poisson distribution. 
Consistent with the expectations for such a non-sharp transition, the behavior of the moments post-transition varies depending on the chosen initial state. For the initial states $\textcolor{mBlue}{|\psi_1\rangle}, \textcolor{mOrange}{|\psi_2\rangle}$ and $\textcolor{mPurple}{|\psi_3\rangle}$, both the LCD moments and the LCD entropy begin to show linear growth.  Conversely, for the other remaining initial states $\textcolor{mGreen}{|\psi_4\rangle}, \textcolor{mPink}{|\psi_5\rangle}, \cdots \textcolor{mRed}{|\psi_7 \rangle}$, both the LCD moments and the LCD entropy initially show a stable plateau before transitioning to linear growth.  The width of the plateau and the slope of the initial growth depend on the initial state. 

\subsubsection{Initial State Dependence}

\begin{figure}[hbt!]
    \centering
    \includegraphics[width=0.95\linewidth]{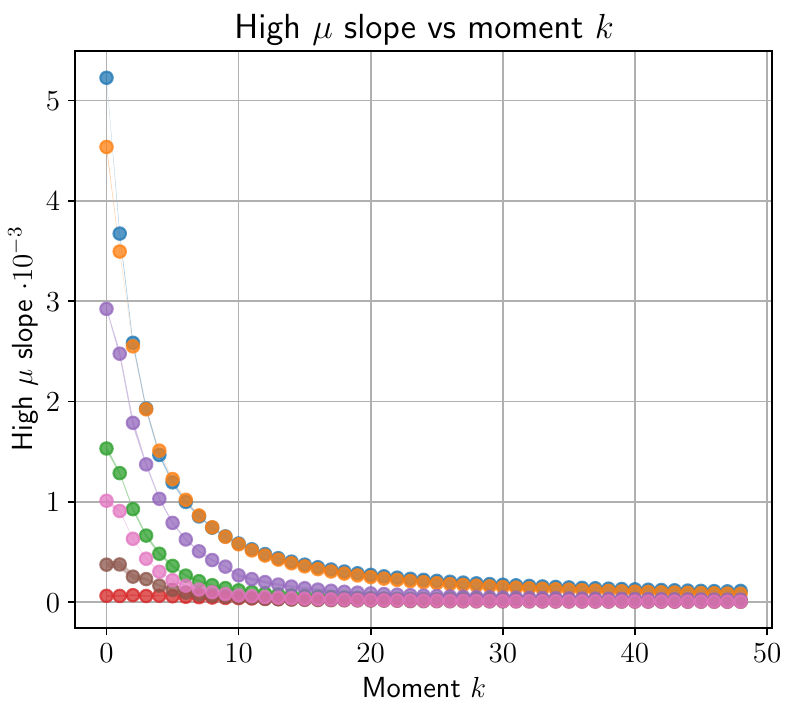}
    \caption{The large $\mu$ slope function as the function of the $k$-th moment. We see that the saturation values depend
    on the initial state.}
    \label{fig:bn_prob_slope_vs_moment}
\end{figure}

To study these dependencies on the initial states qualitatively, we need to relate the width and the slope to a physical observable that characterizes the initial states. A relevant physical observable that characterizes the spreading of a state is given by the Inverse Participation Ratio (IPR), defined as 
\beq
\mathtt{IPR}(\psi) = \sum_{i=1}^N 
\lVert \psi_i \rVert^{4},  
\eeq
where we have assumed that the normalized state $|\psi\rangle$ has $N$ components, each denoted by $\psi_i$. By looking at the large $\mu$ region 
\footnote{A proper definition of the large $\mu$ region is presented in appendix \ref{apd:ErrorAnalysis}.}
, we conclude that the large $\mu$ asymptotic moments are proportional to the IPR of the corresponding initial state, where, for our chosen initial states \eqref{eqn-IntialState}, the IPR is given by the total number of non-vanishing components.
The slope of the region is depicted in Fig. \ref{fig:bn_prob_slope_vs_moment} and we see in the large $k$ limit they all fit into the functional form $\mathtt{c_0}/(k+\mathtt{c}_1) + \mathtt{c}_2$ where the constants $\mathtt{c}_j$ depend on the initial state.

For the LCD entropy, the dependence is more complicated and the endpoint values are not ordered by the IPR. The dominance of the $\textcolor{mOrange}{|\psi_2\rangle}$, which is the only entangled state in all the candidate states, suggests that the entanglement structure of the state also plays a role. For the remaining product states, their endpoint values correlate proportionally with the IPR of the state, similarly to the moments. For results of the R\'{e}nyi entropy for different $\alpha$'s, refer to appendix \ref{apd:RenyiEntropy}.

\subsubsection{Transition Region} \label{sec:transition}

\begin{figure}
    \centering
    \includegraphics[width=\linewidth]{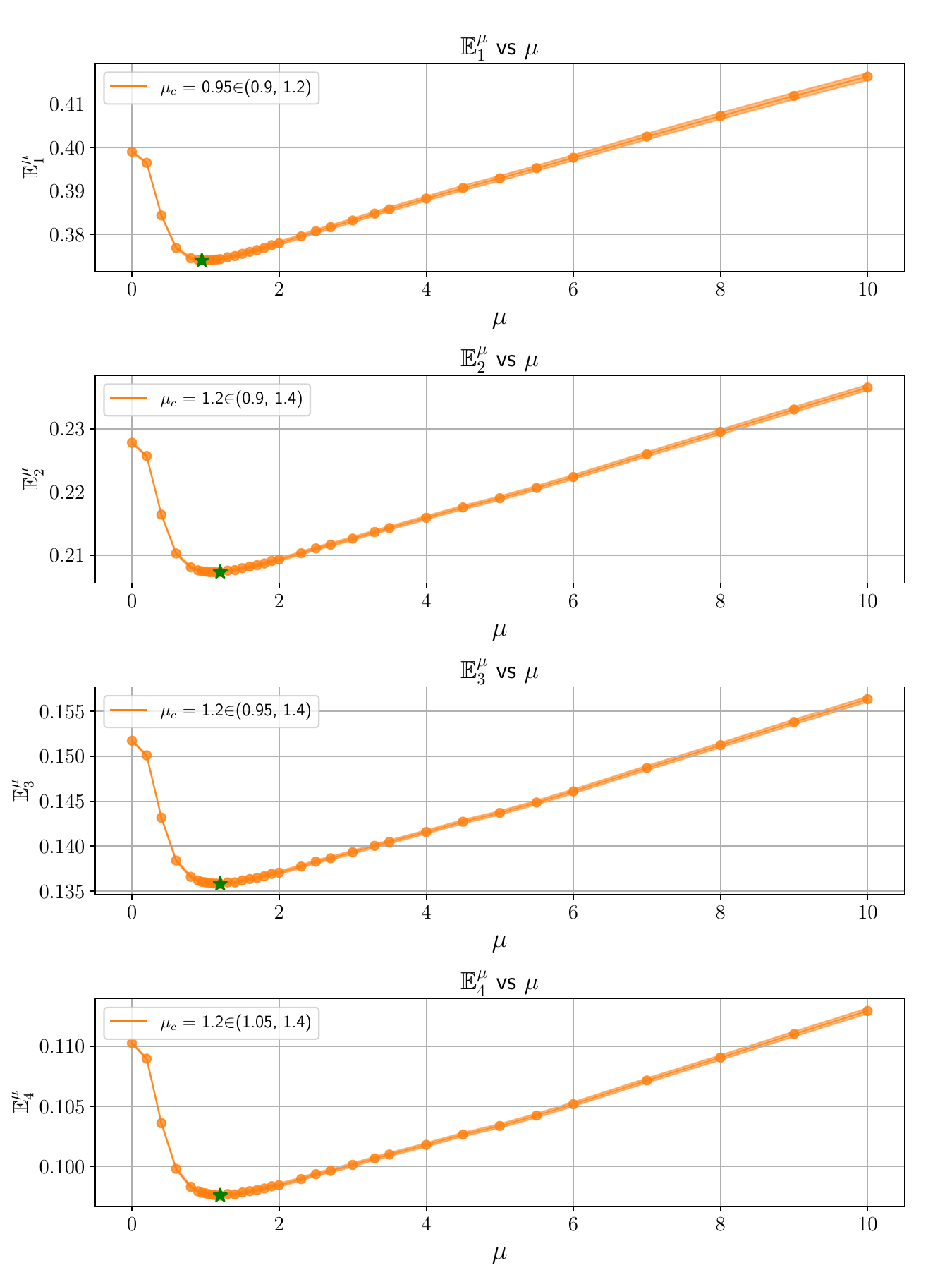}
    \caption{The first four LCD moments as a function of $\mu$ for the state $\textcolor{mOrange}{|\psi_2\rangle}$ defined in \eqref{eqn-IntialState}.}
    \label{fig:bn_states_moments_psi2}
\end{figure}

\begin{figure}
    \centering
    \includegraphics[width=0.95\linewidth]{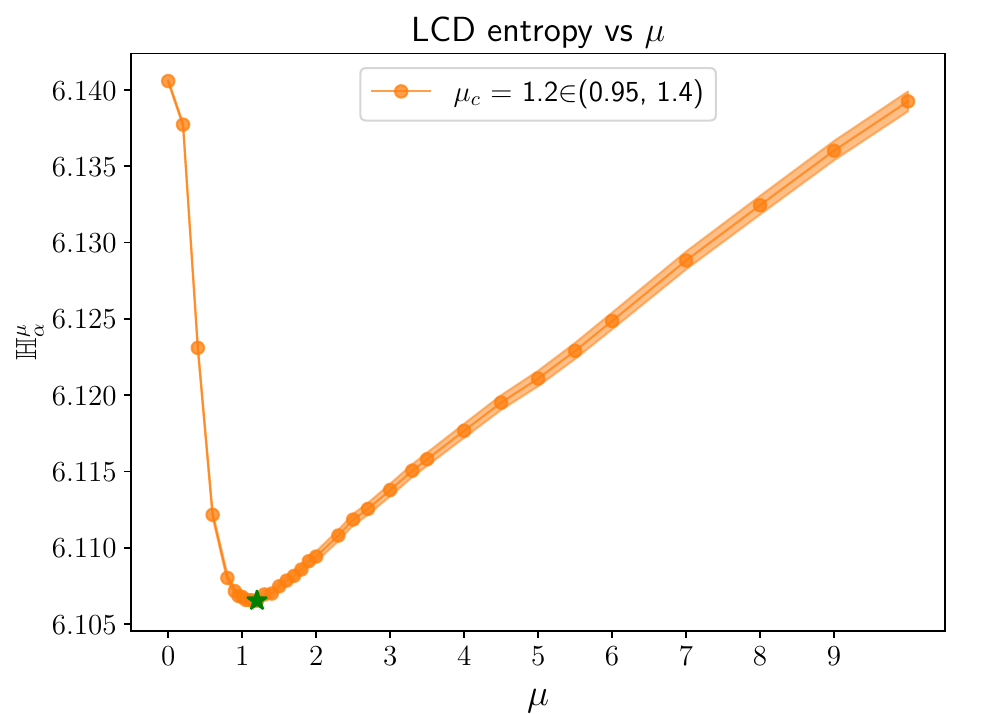}
    \caption{The LCD Shannon entropy as a function of $\mu$ for state $\textcolor{mOrange}{|\psi_2\rangle}$ defined in \eqref{eqn-IntialState}.
    }
    \label{fig:bn_states_entropy_psi2}
\end{figure}

Having understood the state dependence, we now undertake a more detailed examination of the LCD moments. For the sake of simplicity, throughout our analysis, we will focus on the initial state $\textcolor{mOrange}{|\psi_2\rangle}$.
In Fig. \ref{fig:bn_states_moments_psi2} and Fig. \ref{fig:bn_states_entropy_psi2}, we present the first four LCD moments and the Shannon entropy for $\textcolor{mOrange}{|\psi_2\rangle}$, calculated with an enlarged ensemble size of $M=19200$. We observe that an increase in the size of this order ensures the stabilization of numerical errors. Our findings indicate that, except for the first moment, the minima of all moments, as well as the Shannon entropy, are located at $\mu_c \approx 1.2$. 
The minimum of the first moment deviates from this value, but is still within the error bars of $\mu_c \approx 1.2$, as analysed 
in appendix \ref{apd:ErrorAnalysis}. Thus, we cannot decide with certainty, whether the ergodic/MBL transition takes
place as a point, or whether there is a transition region \cite{rao2022power}. 

\begin{figure}
    \centering
    \includegraphics[width=\linewidth]{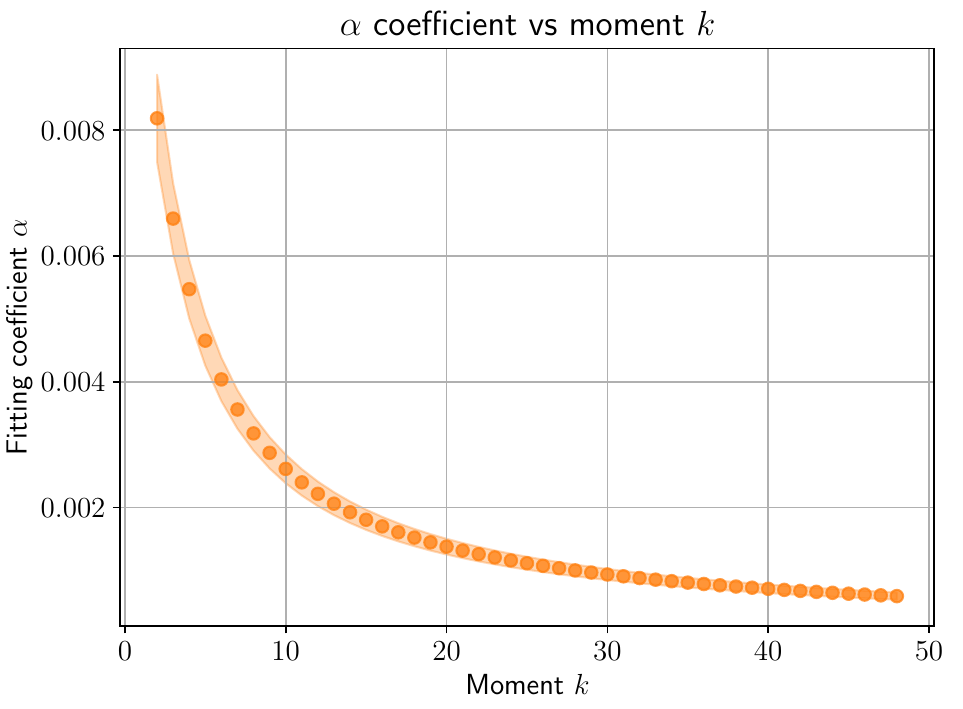}
    \caption{ The coefficient $\alpha_k$ \eqref{scaling} around the critical point $\mu_c \simeq 1.2$ versus different moments. Because of the different location of the minimal point, the first moment was not taken into account in the plot. 
}\label{fig:bn_prob_alpha_vs_moment_psi2}
\end{figure}

For $\textcolor{mOrange}{|\psi_2\rangle}$, as there is no plateau region beyond the transition point, we can fit the scaling of the moments near the phase transition point $\mu_c \simeq 1.2$ as
\begin{equation} \label{scaling}
 \mathbb{E}_k^{\mu} =  \mathbb{E}_k^{\mu= \mu_c} + \alpha_k (\mu-\mu_c)^2 + \mathcal{O}\left( (\mu-\mu_c)^2 \right)\ ,  
\end{equation}
and the result is presented in Fig. \ref{fig:bn_prob_alpha_vs_moment_psi2}. We see that $\alpha_k \sim 1/k $ for $k\gg1$. 
The error in Fig. \ref{fig:bn_prob_alpha_vs_moment_psi2} is smaller for higher moments, which is consistent with our observation that the tail region dominates the higher moments.

\section{Discussion}

The classification of phases of matter requires the introduction of effective diagnostics that distinguish between the different phases.
In this work, we considered the transition between the ergodic and many-body localization phases and 
proposed new diagnostics of the transition, which are based on complexity measures.
We defined the probability distribution function of the Lanczos coefficients of the tri-diagonalized Hamiltonian of the system
and used these complexity measures to analyze the 
power-law random banded matrix model as a function of the correlation strength.

We showed that the moments and the
entropy of the distribution diagnose the ergodic to many-body transition as well as the distinctive feature 
of the phases concerning the memory of the initial conditions. 
The low moments, which are more sensitive to the small $\ell/N$ region, have a memory of the initial state 
feature of the MBL phase, and hence the small $\ell$ Lanczos coefficients
are more effective diagnostics of this.
We found that the LCD distribution has a power-law tail and a compact support, thus the higher moments localize in the same region 
of Lanczos coefficients and exhibit a similar structure.
Our analysis allows within the error bars also the possibility that the ergodic/MBL transition does not happen at a sharp point $\mu=\mu_c$, but rather that there exists a transition region.

Our work opens up a new avenue of using complexity measures in the study of different types of phase transitions at zero and nonzero temperatures, including quantum phase transitions, order/disorder transitions,
confinement/deconfinement phase transition in quantum field theories and holography. 
It would be also interesting to compare our complexity diagnostics to other 
proposed criteria for quantum many-body integrability and chaos, see, e.g., \cite{Khasseh:2023kxw}.

\begin{acknowledgments}
This work is supported in part by the Israeli Science Foundation Excellence Center, the US-Israel Binational Science Foundation, and the Israel Ministry of Science. DlZ is supported in part by the Royal Society University Research Fellowships grant URF/R1/221310 ``Bootstrapping Quantum Gravity." The simulations were conducted on the Imperial College Research Computing Service (DOI: 10.14469/hpc/2232). 
\end{acknowledgments}

\bibliography{MBL}

\appendix
\onecolumngrid
\newpage 

\section*{Supplemental Material}

\section{Analytical Moments for GOE Ensembles}

As a consistency check of our numerical result, in this section, we present some analytical results of the GOE ensemble, corresponding to the $\mu = 0$ case of our model. We will see that the numerical results match perfectly with the analytical results at $\mu =0$ GOE point. 
The Lanczos averages (\ref{b}) define a discrete probability distribution that 
converges in the large $N$ limit to a continuous one. We define in the large $N$ limit a continuous variable $x \equiv \ell/N \in [0,1]$ and denote the limiting continuous distribution as $\mathtt{b}^\mu(x)$. 
When $\mu=0$ we get:
\beq
\mathtt{b}^{\mu = 0}(x) = \frac{3}{2} \sqrt{1-x} +\mathcal{O}(\frac{1}{N})\, . 
\eeq

\begin{figure}[!h]
    \centering
    \includegraphics[width = 0.6 \textwidth]{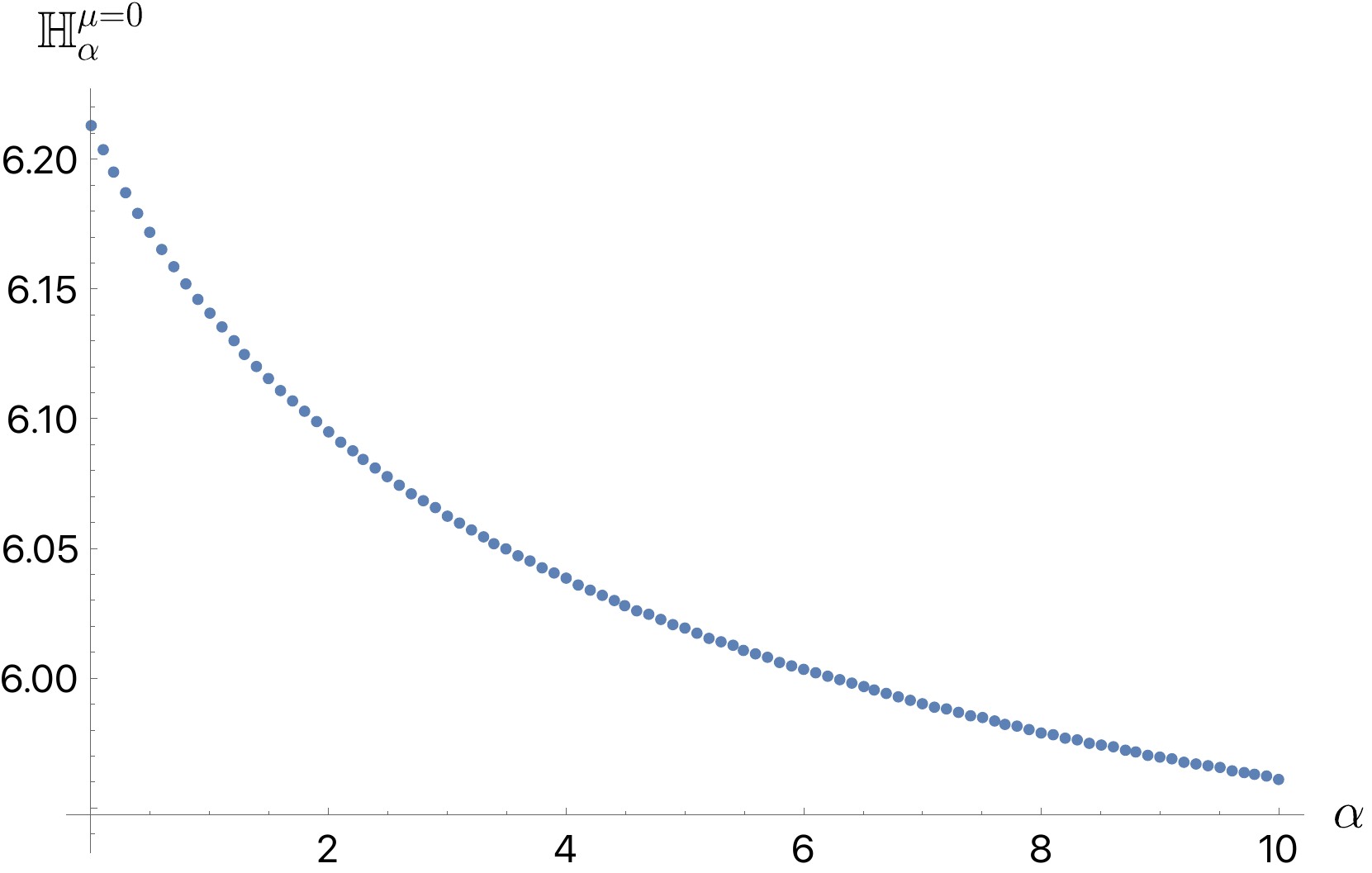}
    \caption{The analytical result of the GOE $\alpha$-R\'{e}nyi entropies with matrix size $N=500$.}
    \label{fig:hEGOE}
\end{figure}

By using this function we can evaluate all the moments. They read
\beq
\mathbb{E}_k^{\mu = 0} \simeq \int_0^1 dx\ x^k\ \mathtt{b}_\ell^{\mu = 0}(x) = \frac{3 \sqrt{\pi } \Gamma (k+1)}{4 \Gamma \left(k+\frac{5}{2}\right)}\, .
\eeq
The first few moments are given by
\beq
\mathbb{E}_1^{\mu = 0} = \frac{2}{5}, \quad \mathbb{E}_2^{\mu = 0} = \frac{8}{35} \simeq 0.23, \quad \mathbb{E}_3^{\mu = 0} = \frac{16}{105} \simeq 0.15, \quad \mathbb{E}_4^{\mu = 0} = \frac{128}{1155} \simeq 0.11,
\eeq
and we find perfect agreement with the numerics at $\mu=0$, see Fig. \ref{fig:bn_states_moments}.
We also compute the R\'{e}nyi entropy using the analytical result \eqref{ab}, see Fig. \ref{fig:hEGOE} for the results. The values match perfectly with the $\mu=0$ values of the entropy plots \ref{fig:bn_states_entropy} and \ref{fig:bn_states_entropy_psi2}.

\section{LCD Plots for All States} \label{apd:LCD}

\begin{figure}[!h]
    \begin{subfigure}[b]{0.32\textwidth}
        \centering
         \includegraphics[width=\textwidth]{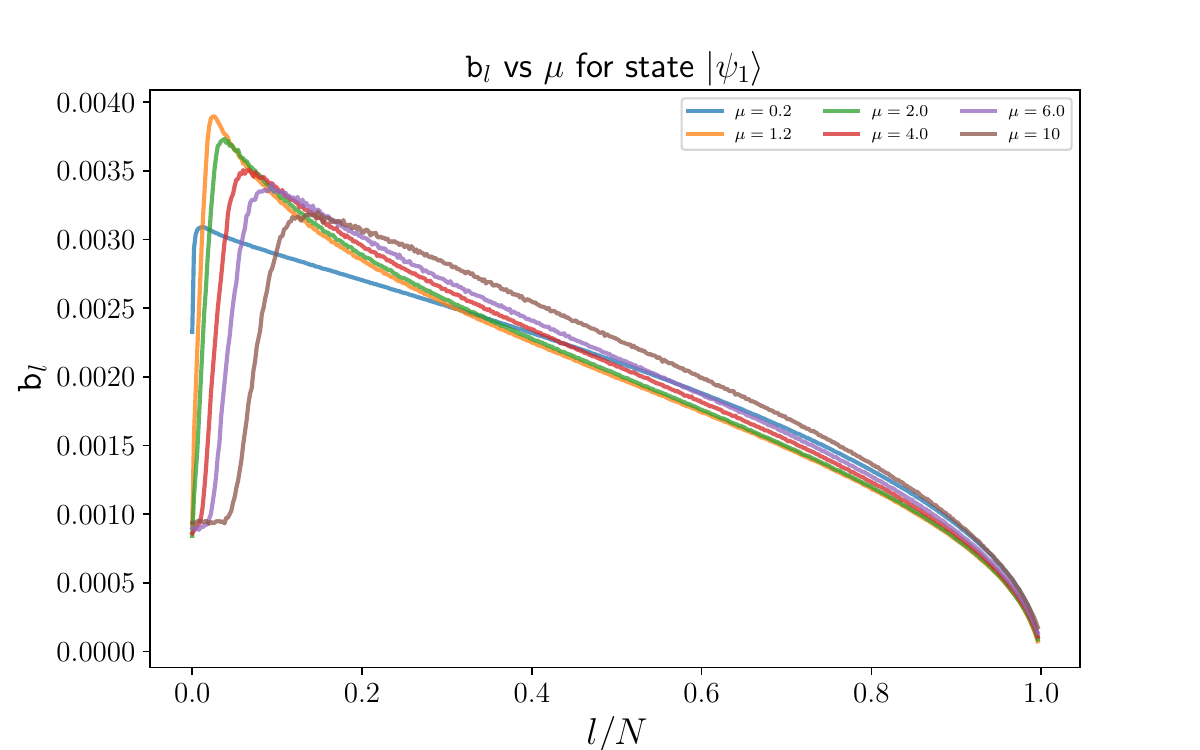}
     \end{subfigure}   
        \hfill
    \begin{subfigure}[b]{0.32\textwidth}
        \centering         \includegraphics[width=\textwidth]{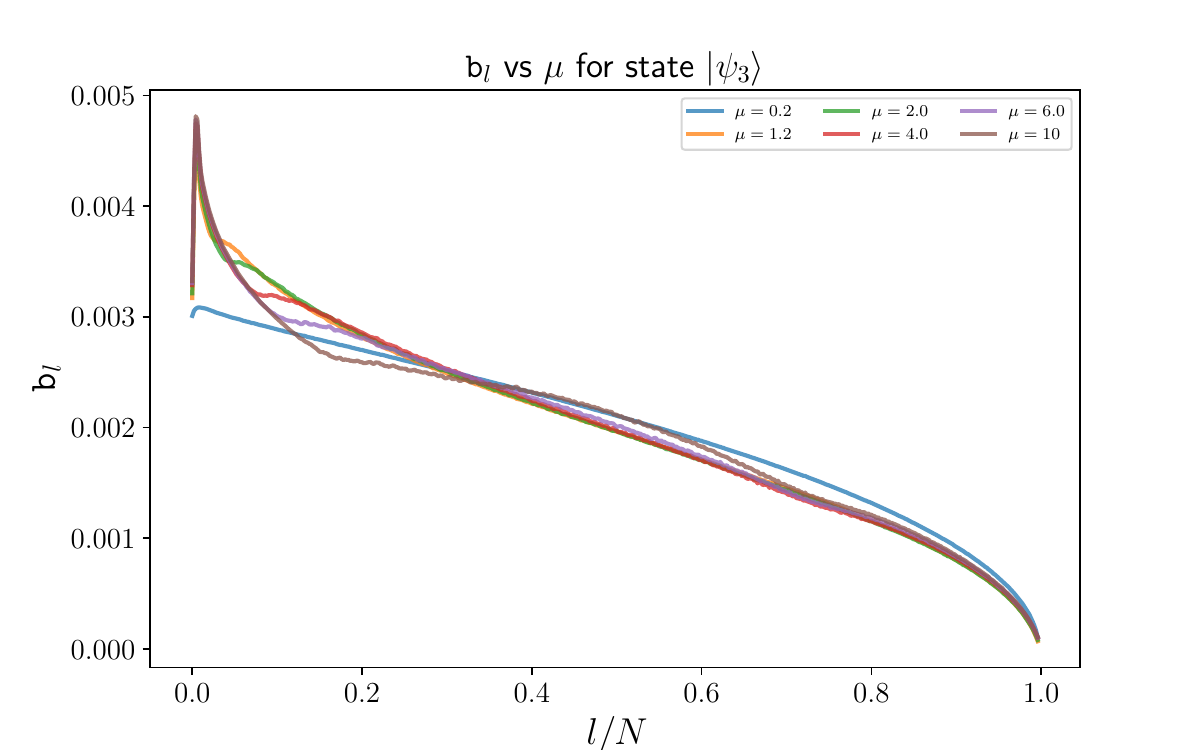}
     \end{subfigure}   
        \hfill
    \begin{subfigure}[b]{0.32\textwidth}
        \centering
         \includegraphics[width=\textwidth]{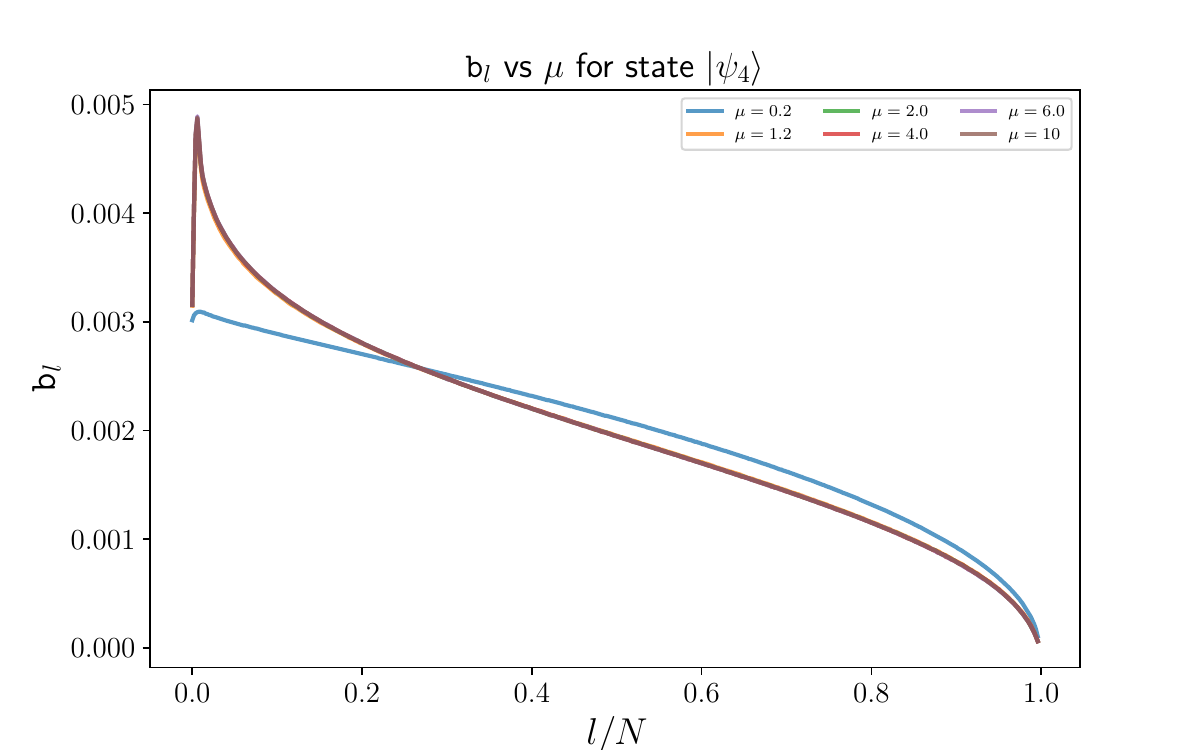}
     \end{subfigure}
     \begin{subfigure}[b]{0.32\textwidth}
        \centering
         \includegraphics[width=\textwidth]{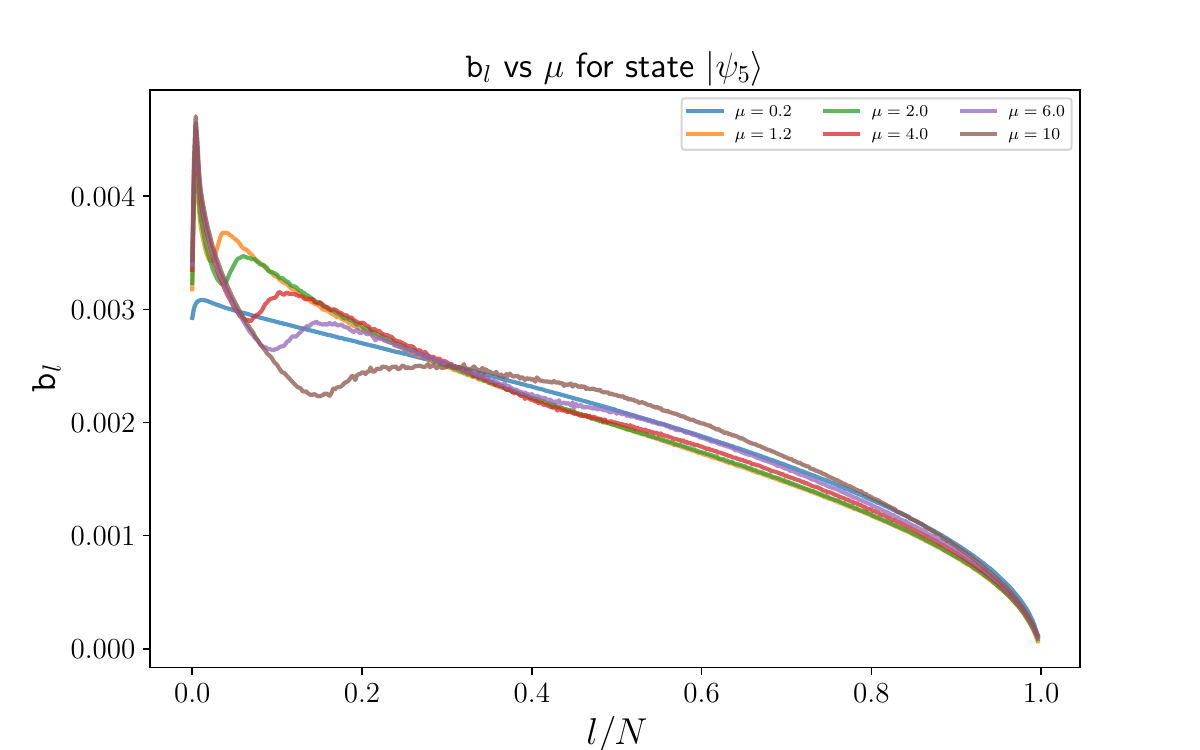}
     \end{subfigure}   
        \hfill
    \begin{subfigure}[b]{0.32\textwidth}
        \centering         \includegraphics[width=\textwidth]{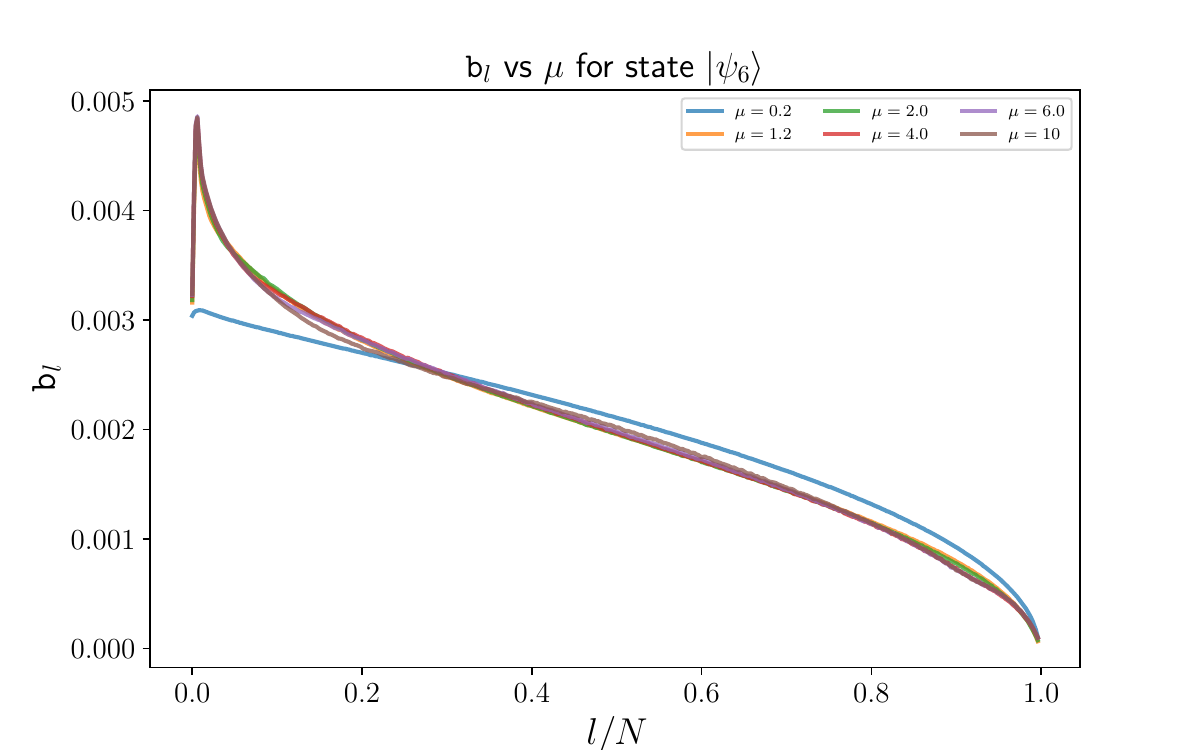}
     \end{subfigure}   
        \hfill
    \begin{subfigure}[b]{0.32\textwidth}
        \centering
         \includegraphics[width=\textwidth]{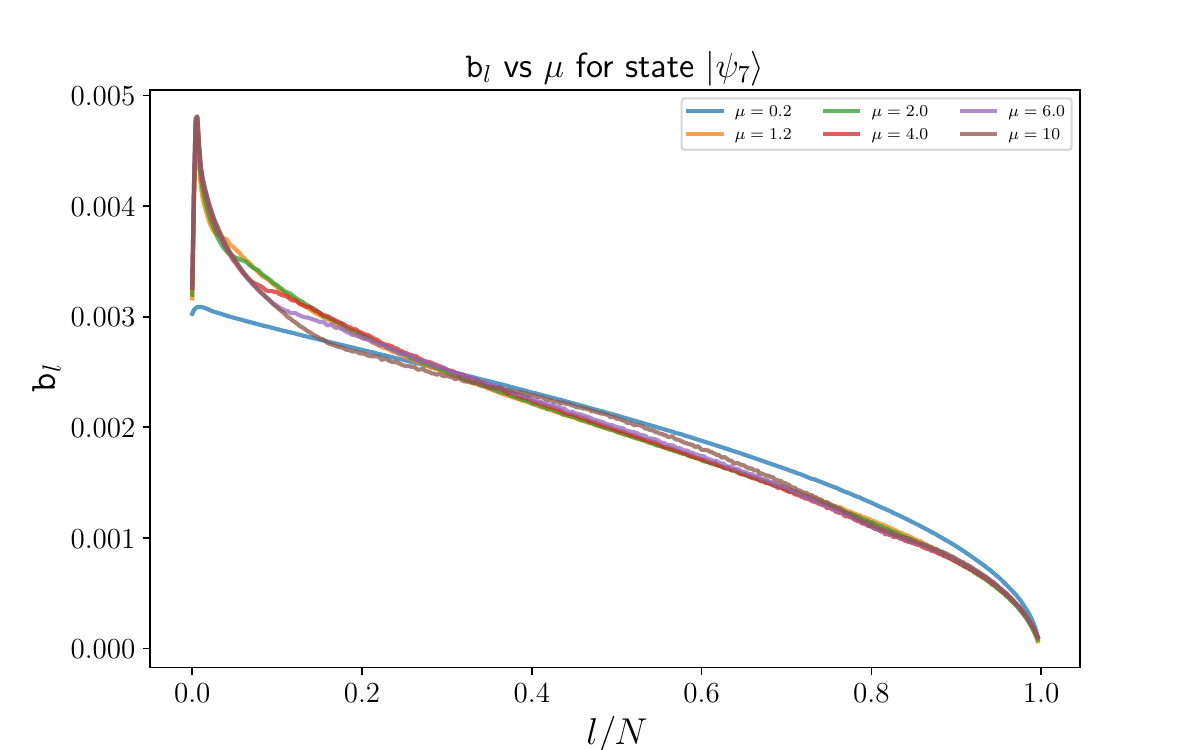}
     \end{subfigure}  
    \caption{LCD for initial states at $\mu = {0.2, 1.2, 2, 4, 6, 10}$: one plot per state.
} \label{fig:LCD_psi_all}
\end{figure}

\begin{figure}
    \begin{subfigure}[b]{0.32\textwidth}
        \centering
         \includegraphics[width=\textwidth]{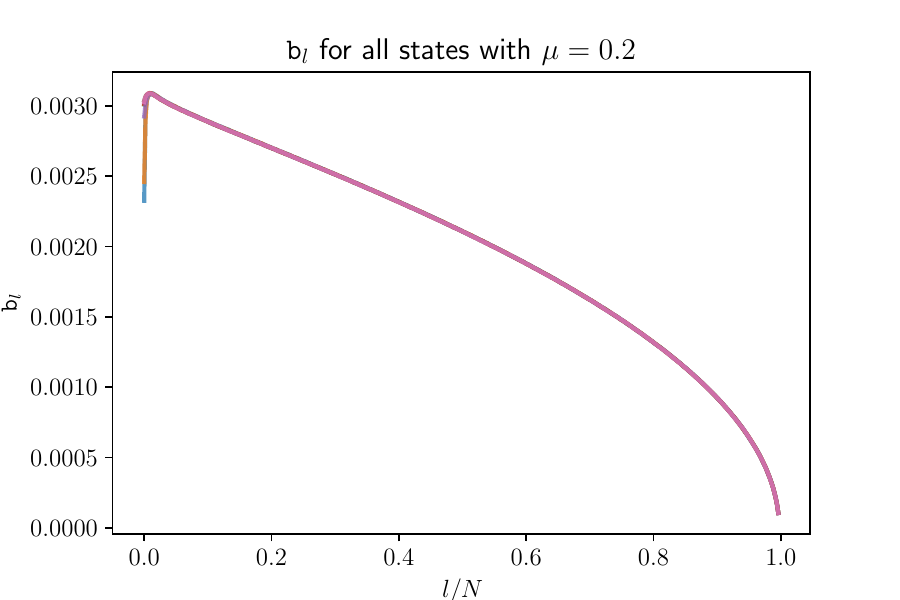}
     \end{subfigure}   
    \begin{subfigure}[b]{0.32\textwidth}
        \centering         \includegraphics[width=\textwidth]{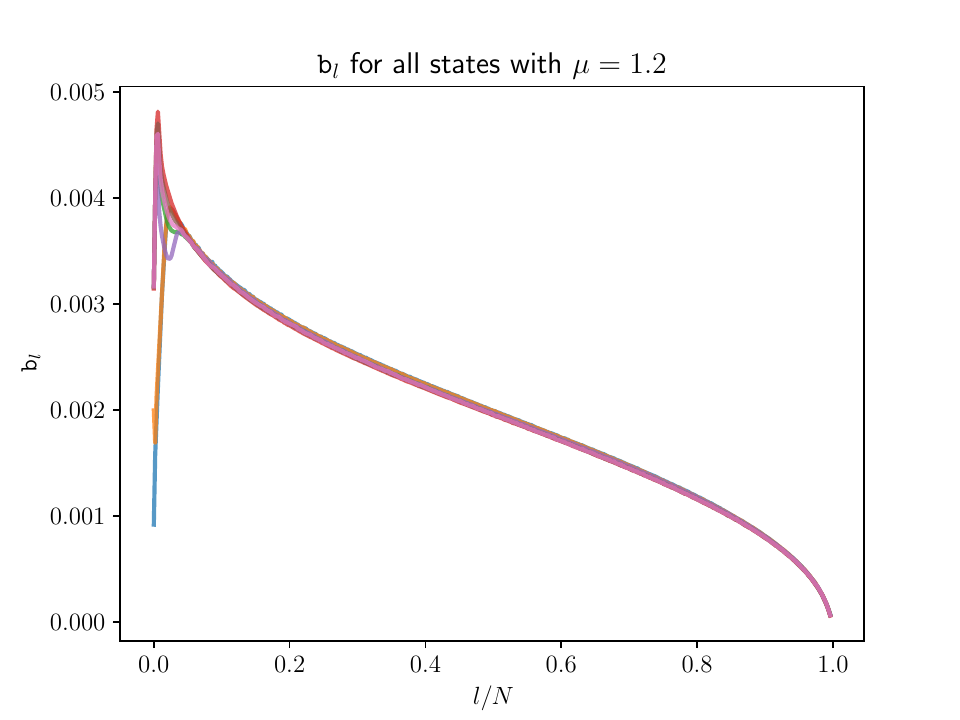}
     \end{subfigure} 
     \begin{subfigure}[b]{0.32\textwidth}
        \centering         \includegraphics[width=\textwidth]{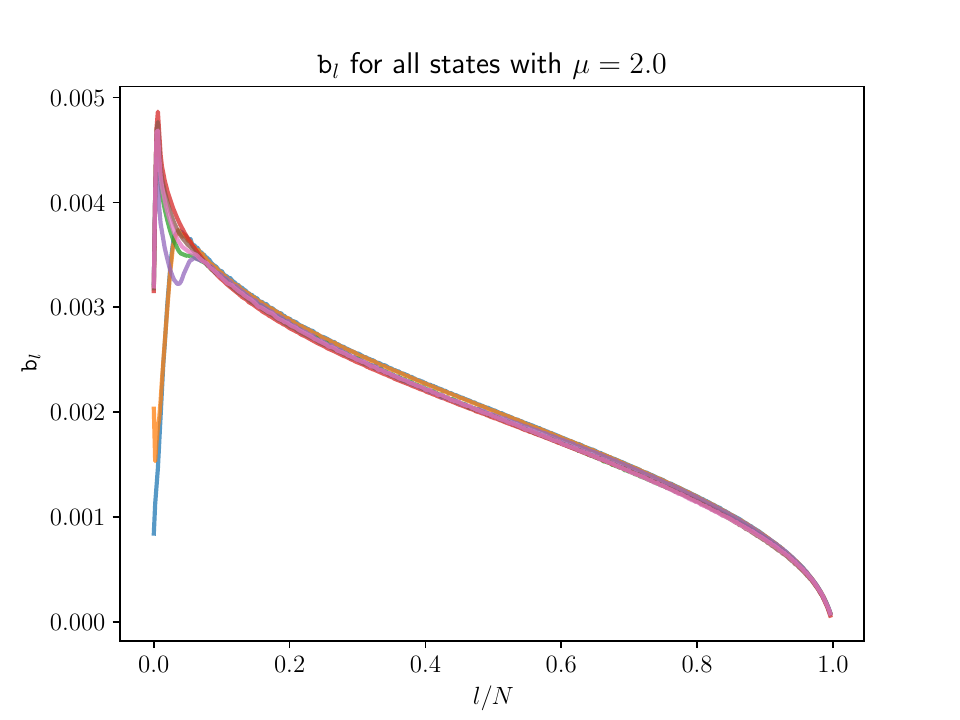}
     \end{subfigure}  
     \begin{subfigure}[b]{0.32\textwidth}
        \centering
         \includegraphics[width=\textwidth]{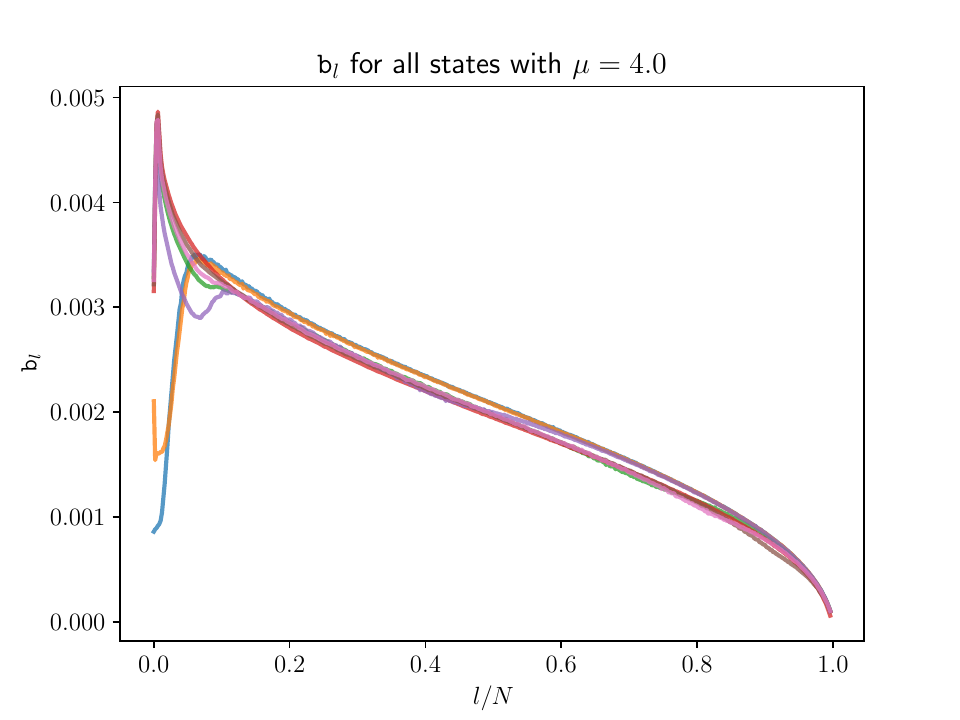}
     \end{subfigure}   
    \begin{subfigure}[b]{0.32\textwidth}
        \centering         \includegraphics[width=\textwidth]{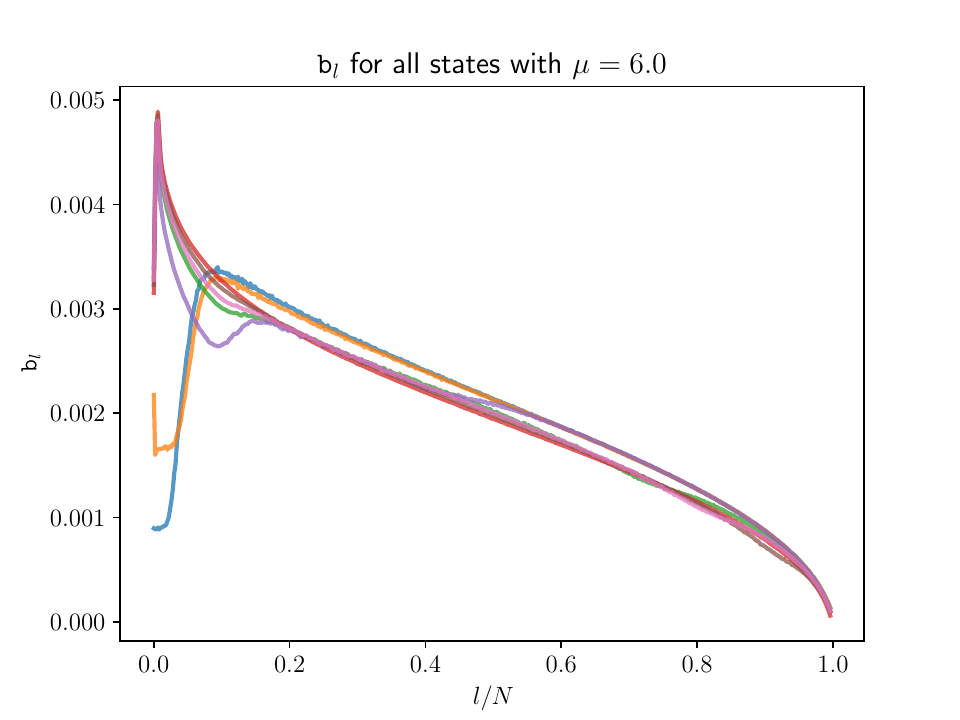}
     \end{subfigure} 
     \begin{subfigure}[b]{0.32\textwidth}
        \centering         \includegraphics[width=\textwidth]{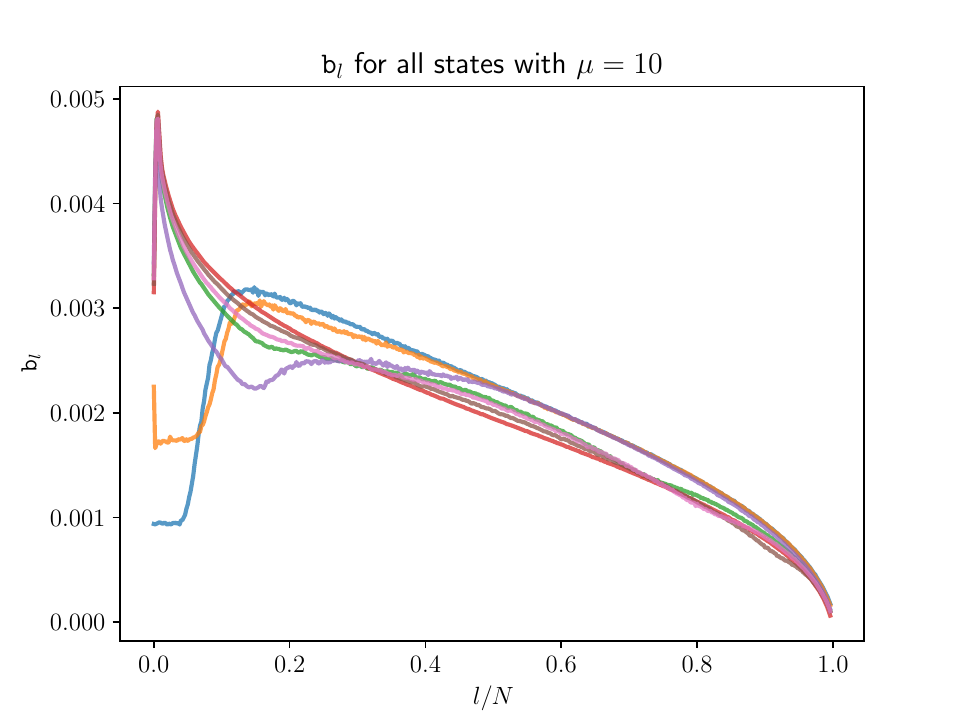}
     \end{subfigure} 
     \caption{LCD for initial states at $\mu = {0.2, 1.2, 2, 4, 6, 10}$: one plot per $\mu$. We see that the main difference is at low $\frac{\ell}{N}$. The tails of the distributions exhibit a power-law.
} \label{fig:LCD_vs_mu}
\end{figure}

In this section, we display the remaining LCD plots for all initial states specified in \eqref{eqn-IntialState}. Fig. \ref{fig:LCD_psi_all} provides a separate plot for each state, with variations in $\mu$ depicted within each plot. 
Fig. \ref{fig:LCD_vs_mu} presents the data differently by providing separate plots for each $\mu$. This allows for the visualization of differences among states within each plot. The qualitative result in the section \ref{sec:LCD} remains the same.

\newpage
\section{LCD Moments Summand Plot}

\begin{figure}[!h]
    \begin{subfigure}[b]{0.32\textwidth}
        \centering
         \includegraphics[width=\textwidth]{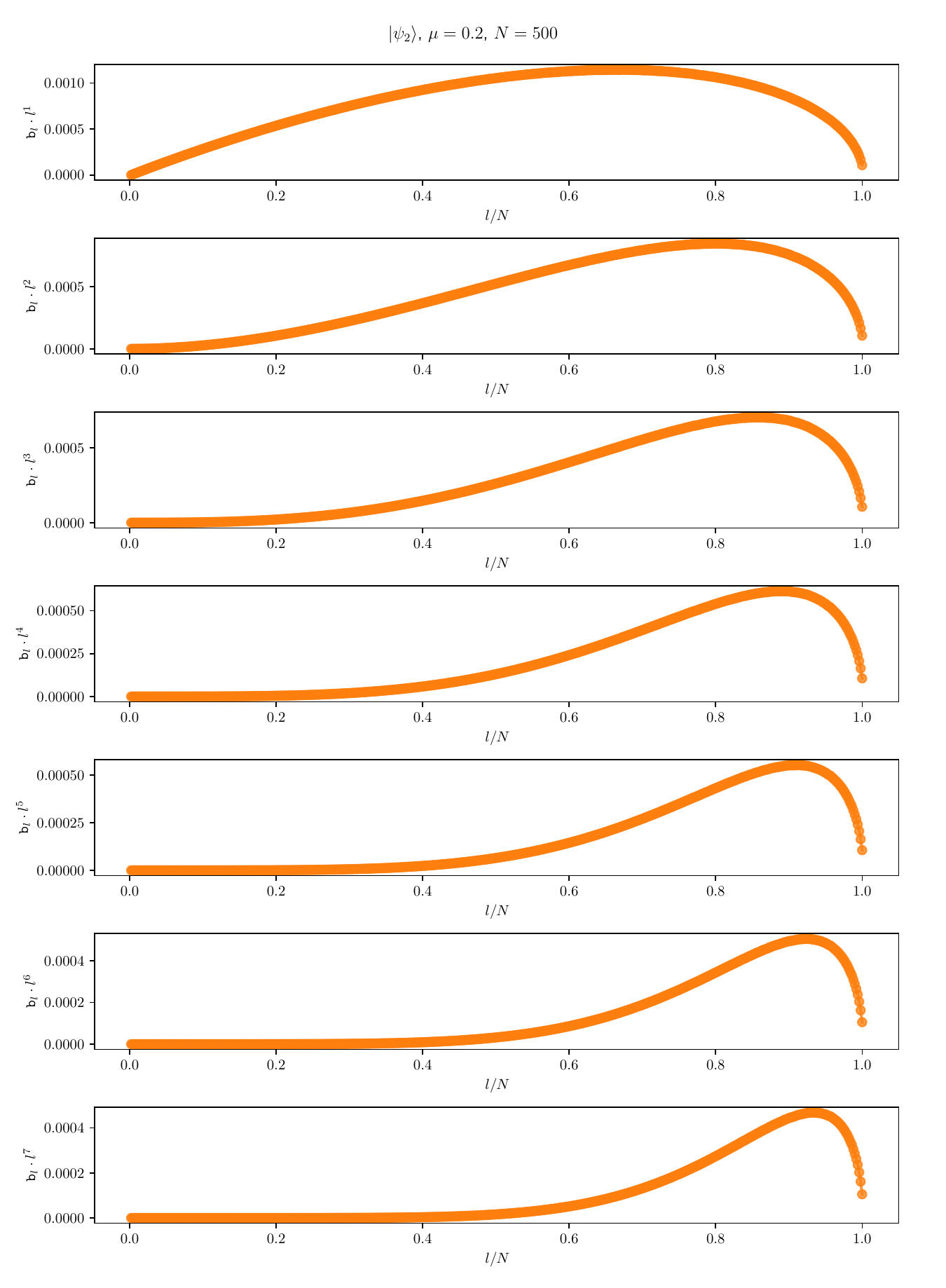}
     \end{subfigure}   
    \begin{subfigure}[b]{0.32\textwidth}
        \centering         \includegraphics[width=\textwidth]{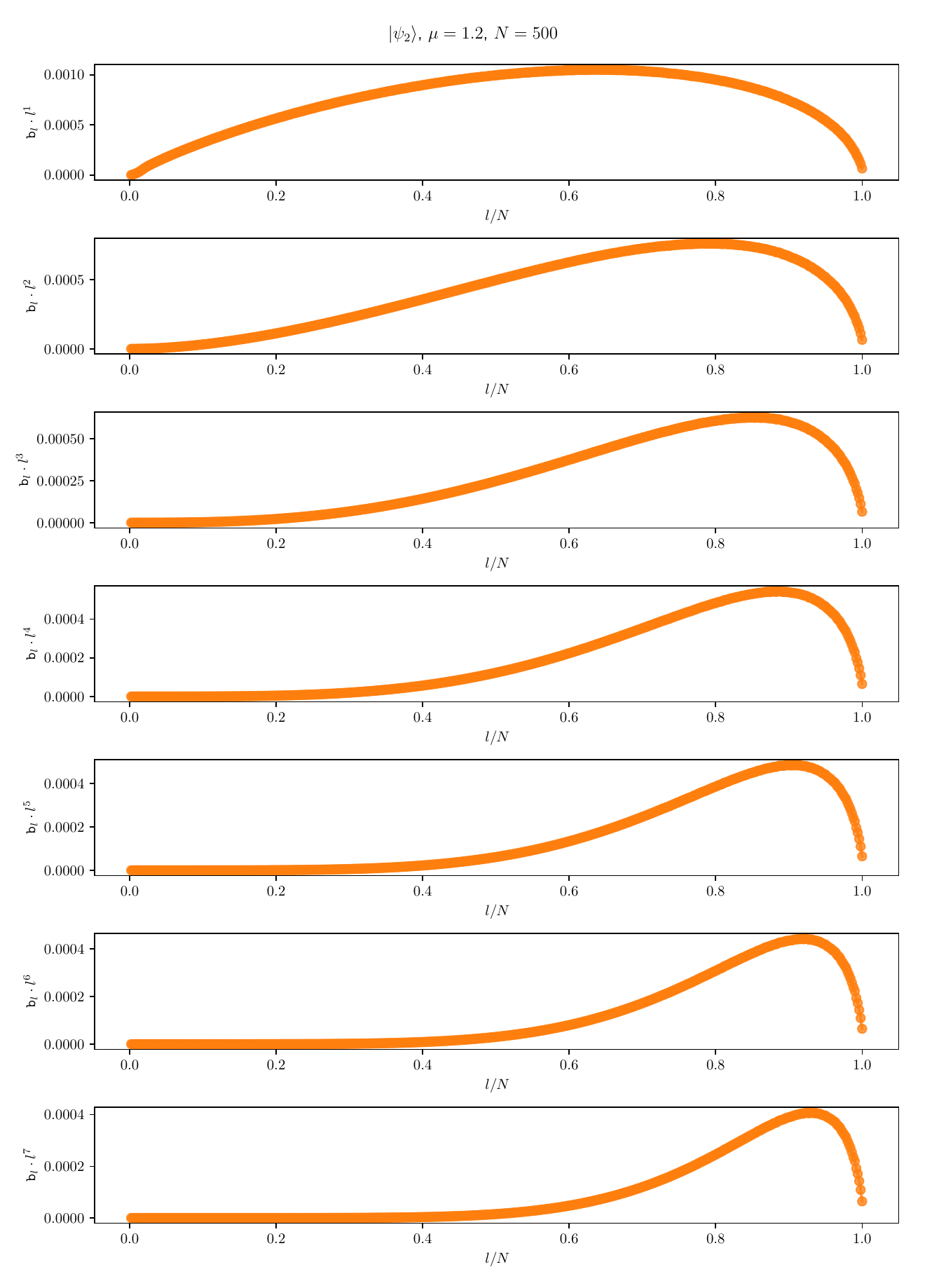}
     \end{subfigure} 
     \begin{subfigure}[b]{0.32\textwidth}
        \centering         \includegraphics[width=\textwidth]{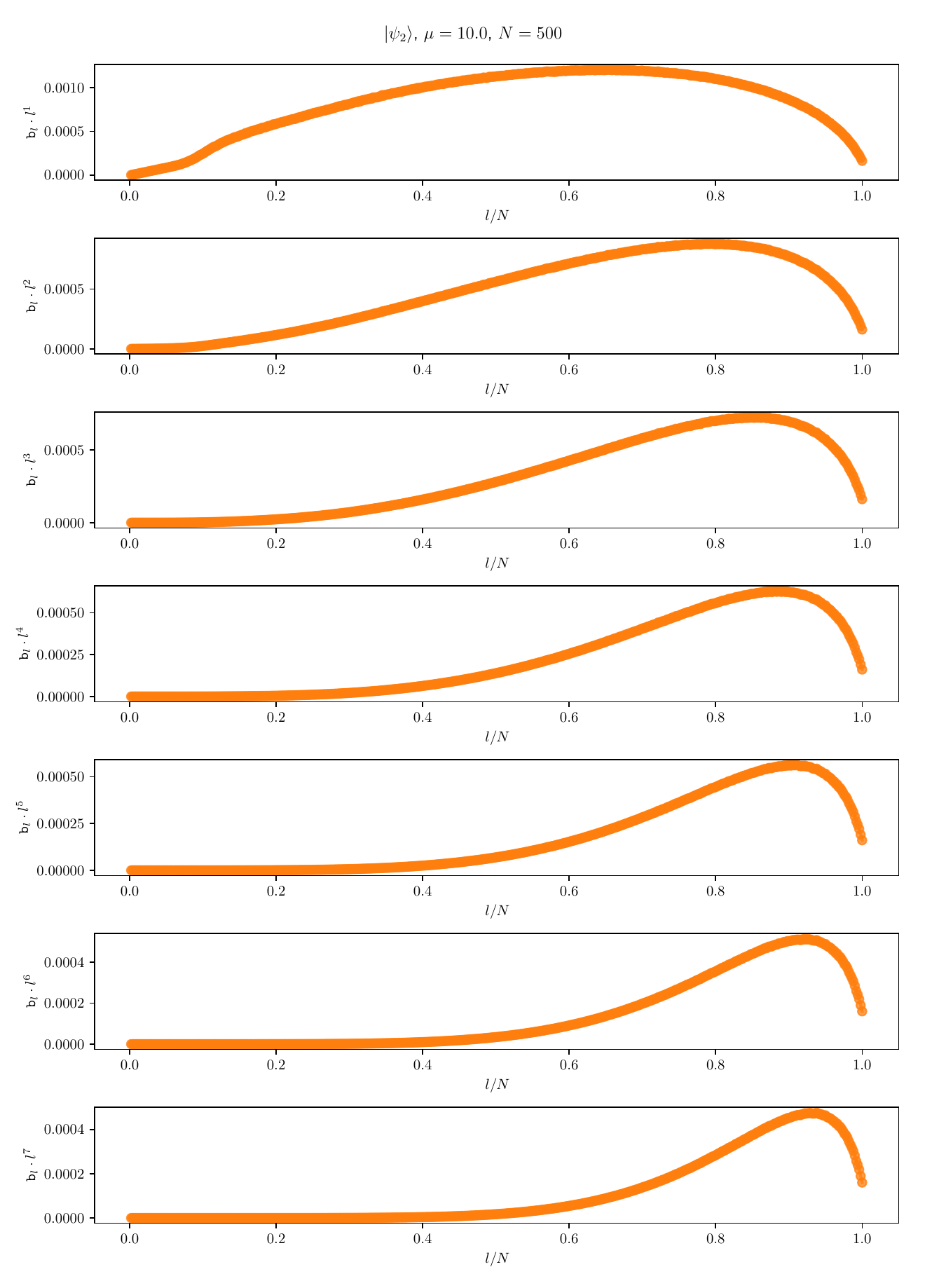}
     \end{subfigure}  
     \caption{$\mathtt{b}_l*l^k$ for $k=1,\cdots, 8$ for $\textcolor{mOrange}{|\psi_2\rangle}$ state with different $\mu$ values.  We see
     that the main contribition for the high moments is from the same high $\frac{\ell}{N}$ region.
} \label{fig:LCD_moments_raw_psi2}
\end{figure}

\begin{figure}[!h]
    \begin{subfigure}[b]{0.32\textwidth}
        \centering
     \includegraphics[width=\textwidth]{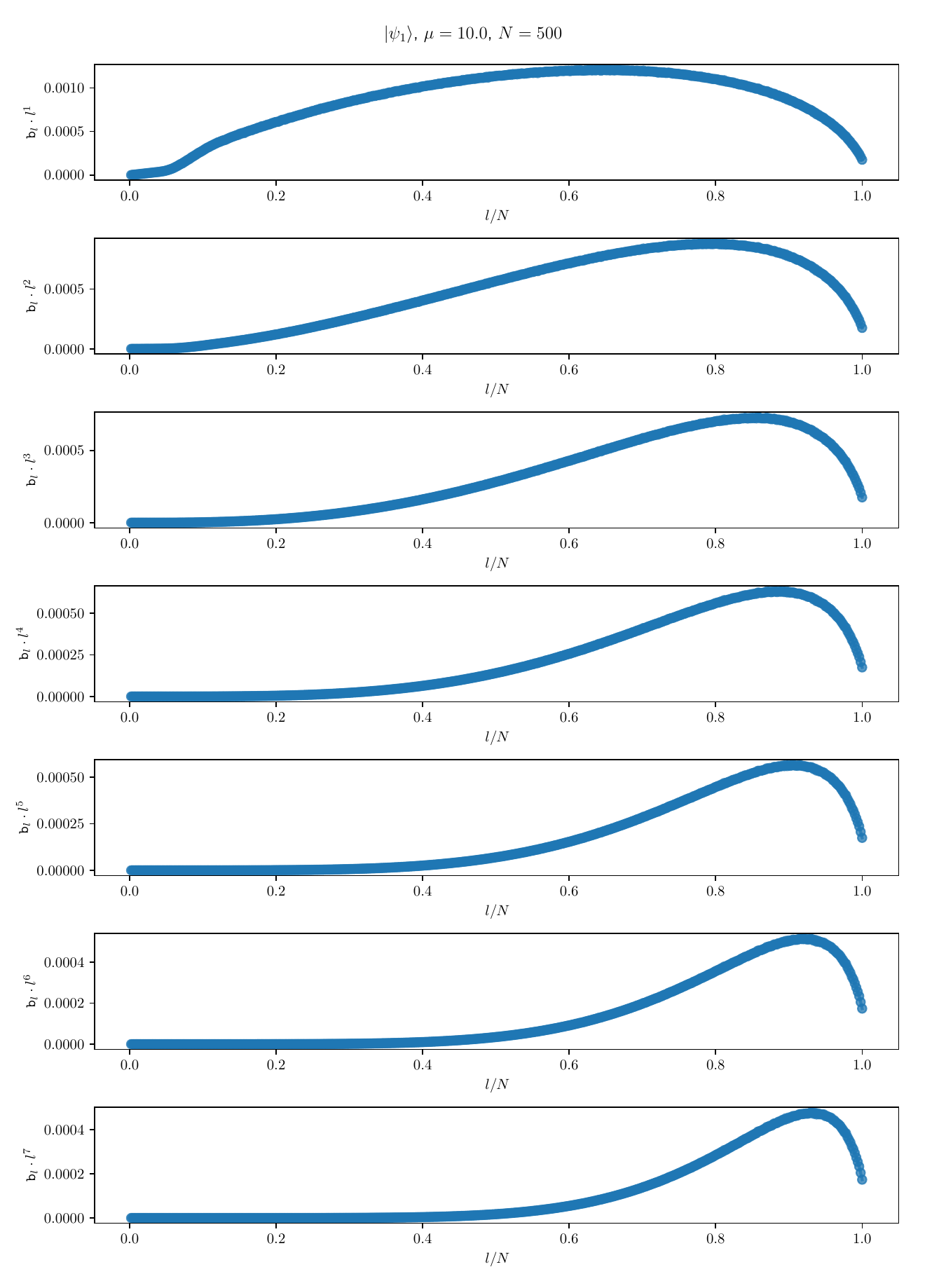}
     \end{subfigure}   
    \begin{subfigure}[b]{0.32\textwidth}
        \centering         \includegraphics[width=\textwidth]{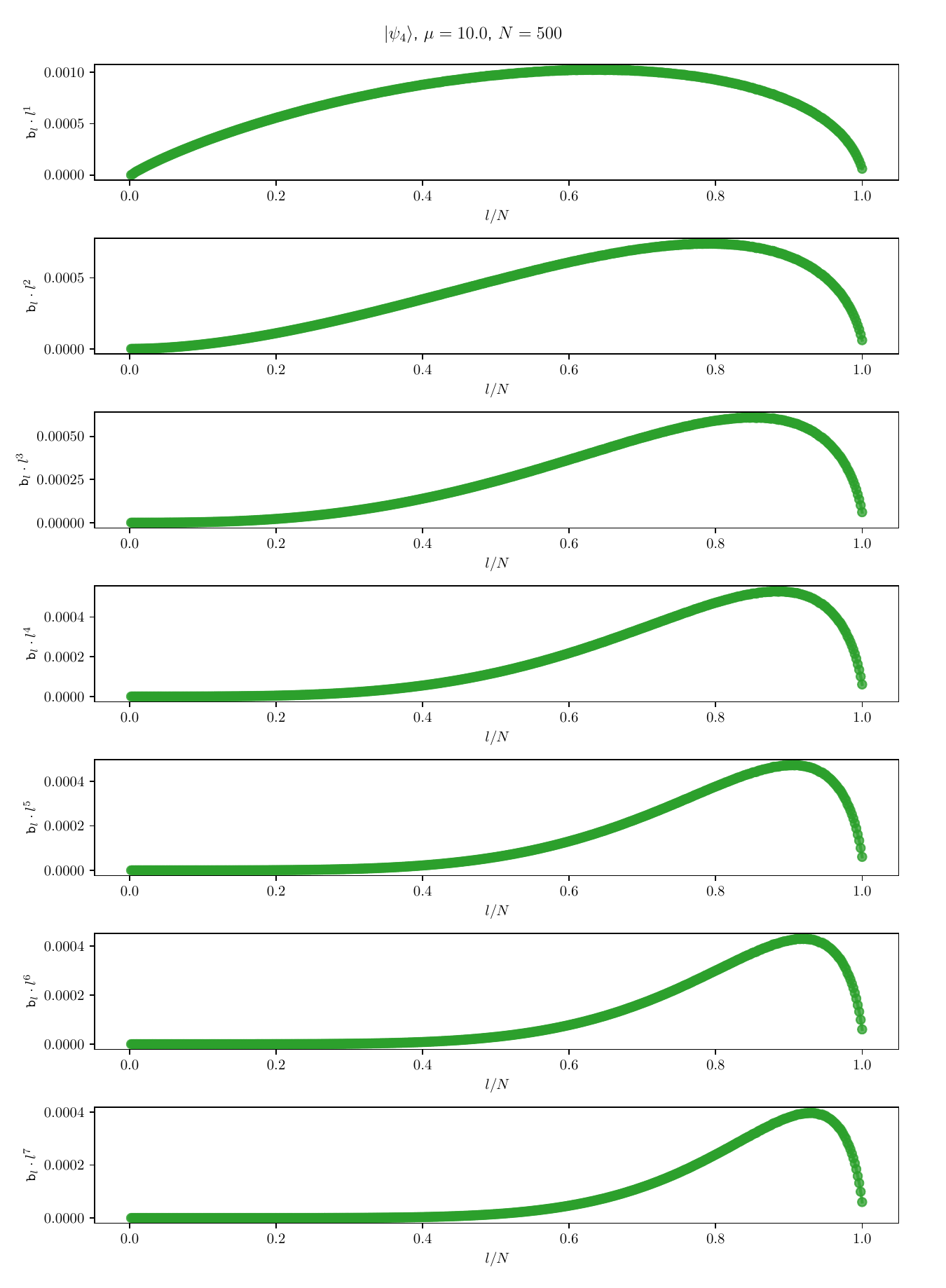}
     \end{subfigure} 
     \begin{subfigure}[b]{0.32\textwidth}
        \centering         \includegraphics[width=\textwidth]{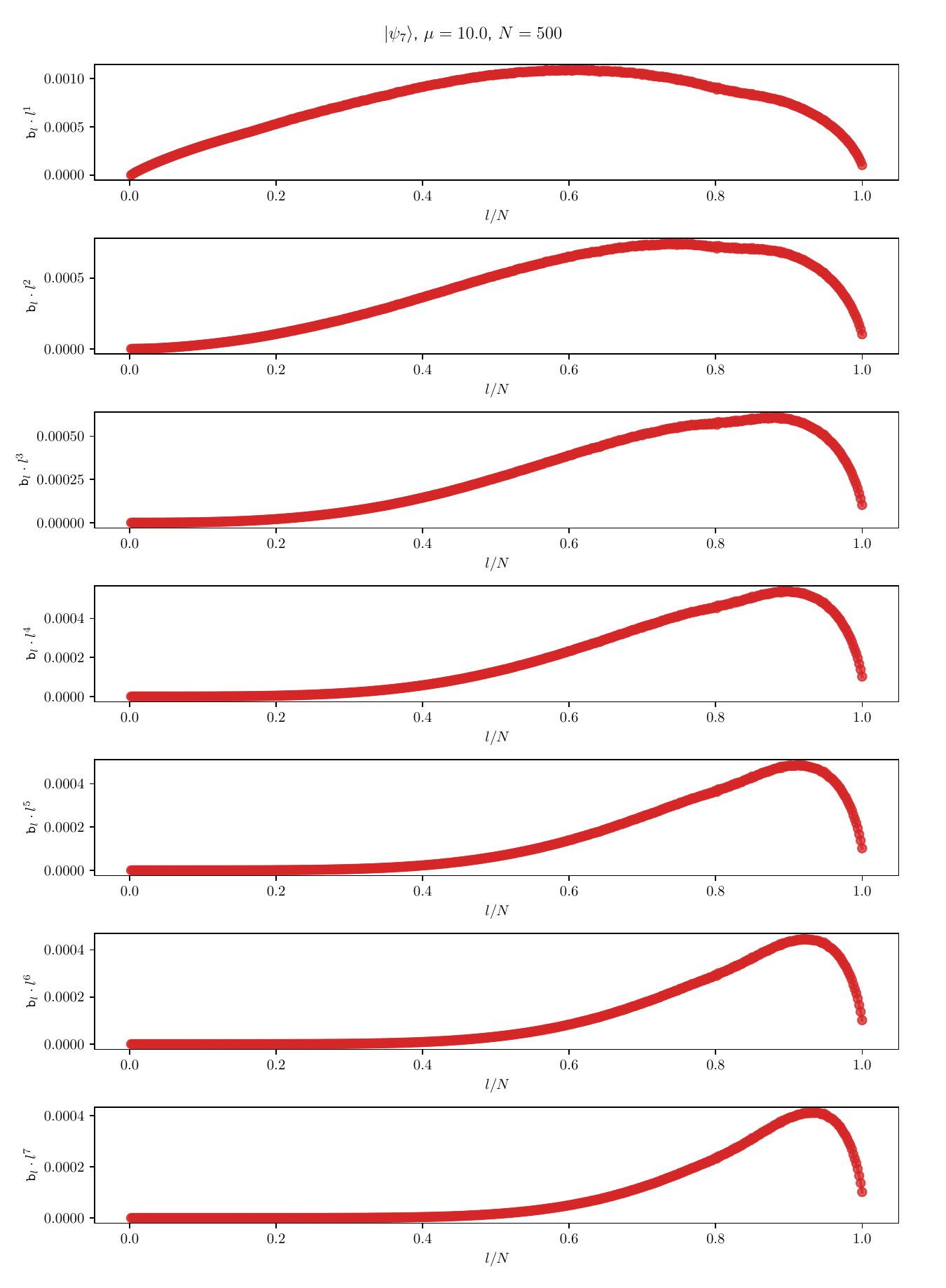}
     \end{subfigure}  
     \caption{The summands of the $k$-moment $\mathtt{b}_{\ell}*\left(\frac{\ell}{N}\right)^k$ for $k=1,\cdots, 8$ for initial states $\textcolor{mBlue}{|\psi_1\rangle}$, $\textcolor{mGreen}{|\psi_4\rangle}$, and $\textcolor{mRed}{|\psi_7\rangle}$ with $\mu = 10$. We see
     that the main contribution for the high moments is from the same high $\frac{\ell}{N}$ region.
} \label{fig:LCD_moments_raw_all}
\end{figure}

To clarify the similarities between the high moments for different states, we plotted the function $\mathtt{b}_\ell \times (\ell/N)^k$ for values of $k$ ranging from one to eight, and for different initial states in Figs. \ref{fig:LCD_moments_raw_psi2} and \ref{fig:LCD_moments_raw_all}. Recall that the LCD is a probability distribution with a finite range of support and has a power-law tail, thus mostly dominated by the small $\ell/N$ values. Conversely, when computing higher moments, the factor $(\ell/N)^k$ becomes large for large $\ell$ values, resulting in a competition between the two effects. The data shows that the major contributions to higher moments mostly come from the tail of the LCD where the ratio of  $\ell/N \sim 1$. 
Considering that the LCD has almost no dependence on the state within this range, it can be inferred that the higher moments in different states are similar.

\section{The large $\mu$ Ensemble} \label{apd:Poisson}

\begin{figure}[hbt!]
    \centering
    \includegraphics[width=0.6\linewidth]{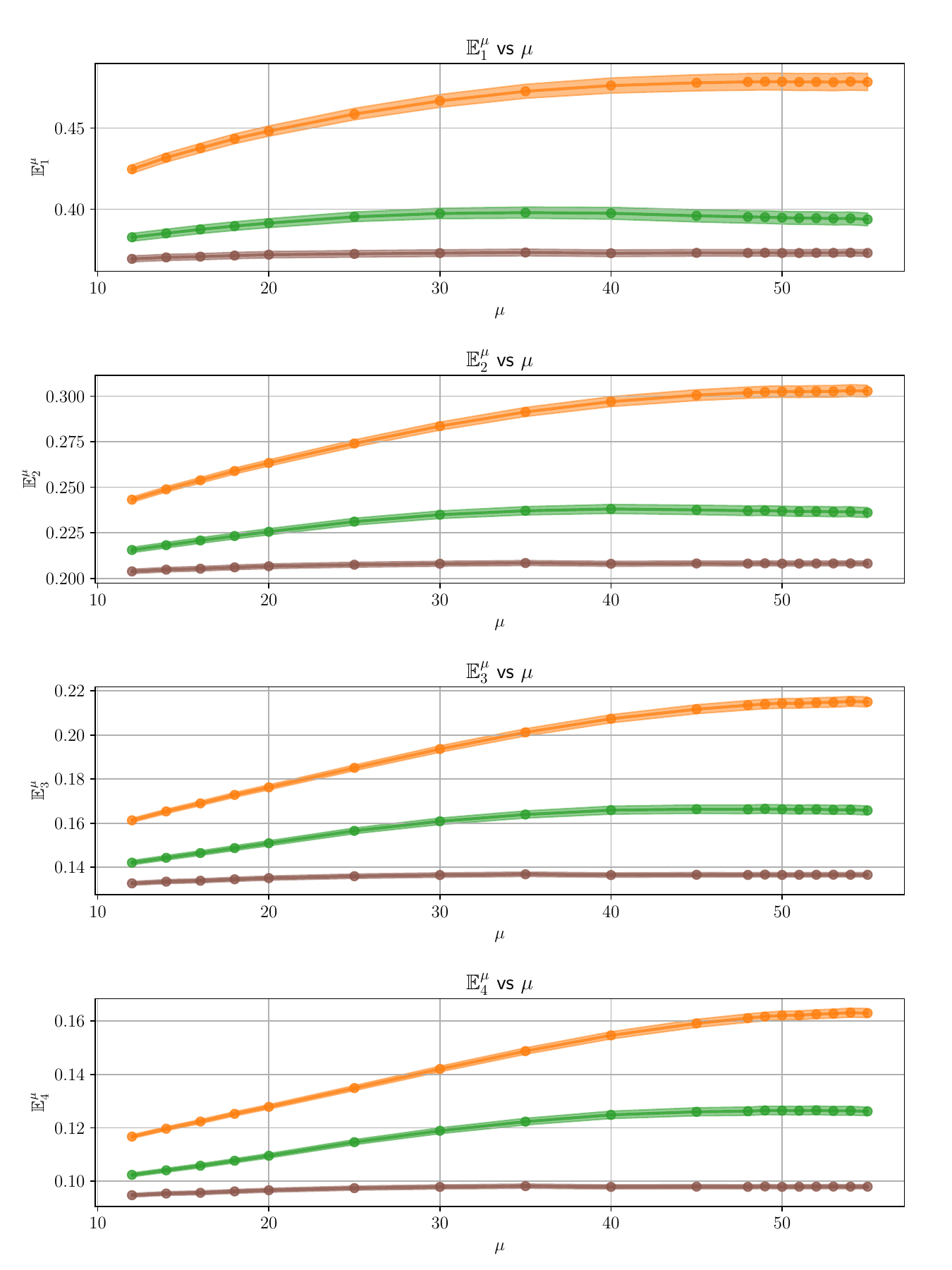}
    \caption{The first four LCD moments as functions of $\mu \in \{ 12, 14, 16, 18, 20, 25, 30, 35, 40, 45, 48, 49, 50, 51, 52, 53, 54, 55 \}$ for initial states $\textcolor{mOrange}{|\psi_2\rangle}$, $\textcolor{mGreen}{|\psi_4\rangle}$ and $\textcolor{mBrown}{|\psi_5\rangle}$ with error bars.
    The matrix size is $N=500$ and the ensemble size is $M=2400$. We observe saturation around $\mu \simeq \sqrt{M} \simeq 49$.}
    \label{fig:moments_largeMu}
\end{figure}

The limit $\mu \rightarrow \infty$ corresponds to a Poisson process associated with integrable systems. One expects the Hamiltonian in this case to have a diagonal form with vanishing Lanczos coefficients $b_l$, and hence an ill-defined 
distribution of their averages. In the numerical analysis one observes a saturation of the moments at large $\mu$, see Fig. \ref{fig:moments_largeMu}, which corresponds to a particular order of limits that yields a uniform distribution. In the region where the distributions become uniform, their endpoint values are still ordered according to the corresponding IPR value of the initial states, same as in Fig. \ref{fig:bn_states_moments}.

Consider a finite number of samples $M$.
Taking $\mu \geq \sqrt{M}$ in \eqref{Cor} leads to $\rho = 1$ for $i=j$ and $\rho=\frac{1}{2}$ for $i=j\pm1$ (for $\mathtt{B}=1$). The Hamiltonian then takes the tri-diagonal form (equation \ref{H}): $H_{i,j} = \left( \delta_{i,j} + \frac{1}{2} \delta_{i,j+1} + \frac{1}{2}\delta_{i+1,j} \right) a_{i,j}$, and
we can read the averages of the Lanczos coefficients:  $\langle b^{\mu\rightarrow \infty}_\ell \rangle = \frac{1}{2}\langle a_{i,j} \rangle$, where $ \langle a_{i,j} \rangle \sim \mathcal{N} \left( 0, \frac{1}{\sqrt{M}} \right)$.
The normalized coefficients are given by: 
\begin{gather}
    \mathtt{b}^{\infty} _\ell = \frac{\langle b^\mu_\ell \rangle}{\sum_{\ell=1}^{N-1} \langle b^\mu_\ell \rangle} = \frac{\langle a_{i,j} \rangle}{\sum_{\ell=1}^{N-1} \langle a_{i,j} \rangle} = \frac{1}{N-1}\, ,
\end{gather}
and we used the fact that all $a_{i,j}$ are independent and identically distributed.

\section{Error Analysis} \label{apd:ErrorAnalysis}

In this section, we recall some formulas that we use for error analysis. For an ensemble of size $M$, the standard deviation is defined as
\beq
\sigma = \sqrt{\frac{\sum_{i=1}^M (x_i - \langle x_i\rangle)}{M-1}}\, .
\eeq
Here, $x_i$ represents individual data points, and $\langle x_i\rangle$ denotes the mean of these data points. Additionally, the \textit{standard error} for the data is defined as the ratio of the standard deviation to the square root of the ensemble size $\delta x \equiv \sigma/\sqrt{M}$.

For independent variables $x_i$ we have the following formula for combining the error,
\begin{itemize}
    \item Sum: if $Q = \sum_i x_i$ then $\delta Q = \sqrt{\sum_I (\delta x_i)^2}$.
    \item Ratio: if $Q = \frac{\sum_i x_i}{\sum_i y_i}$, we have
    \beq
    \frac{\delta Q}{|Q|} = \sqrt{\sum_i \left(\frac{\delta x_i}{x_i}\right)^2 + \sum_i \left(\frac{\delta y_i}{y_i}\right)^2 }\, .
    \eeq
    Our distribution $\mathtt{b}^\mu_\ell$ defined in \eqref{b} is the composition of the two. We then conclude that
    \beq \label{eqn-apd-ErrorB}
    \frac{\delta \mathtt{b}^\mu_\ell}{|\mathtt{b}^\mu_\ell|} = \sqrt{\left(\frac{\delta b^\mu_\ell}{b^\mu_\ell}\right)^2 + \frac{\sum_\ell \left(\delta b^\mu_\ell\right)^2}{(\sum_\ell b^\mu_\ell)^2} }\, .
    \eeq
    \item 
    
    For non-linear functions such as the R\'{e}nyi entropy \eqref{eqn-RenyiE}, we have to use the error propagation formula. It reads,
    \beq
    \delta \mathbb{H}_\alpha^\mu = \sqrt{\sum_\ell \left(\frac{\partial \mathbb{H}_\alpha^\mu}{\partial \mathtt{b}_\ell^\mu} \delta \mathtt{b}_\ell^\mu \right)^2} = \sqrt{\sum_\ell \left( \frac{\alpha}{1-\alpha} \frac{[\mathtt{b}_\ell^\mu]^{\alpha -1}}{\sum_{m} [\mathtt{b}^{\mu}_m]^\alpha } \delta \mathtt{b}_\ell^\mu \right)^2}\, .
    \eeq
    In the $\alpha =1$ case it implies for Shannon entropy,
    \beq
    \delta \mathbb{H}^\mu = \sqrt{\sum_\ell \big( [1+\log \mathtt{b}_\ell^\mu] \delta \mathtt{b}_\ell^\mu \big)^2}\, .
    \eeq
\end{itemize}

\begin{figure}[!h]
    \begin{subfigure}[b]{0.32\textwidth}
        \centering
         \includegraphics[width=\textwidth]{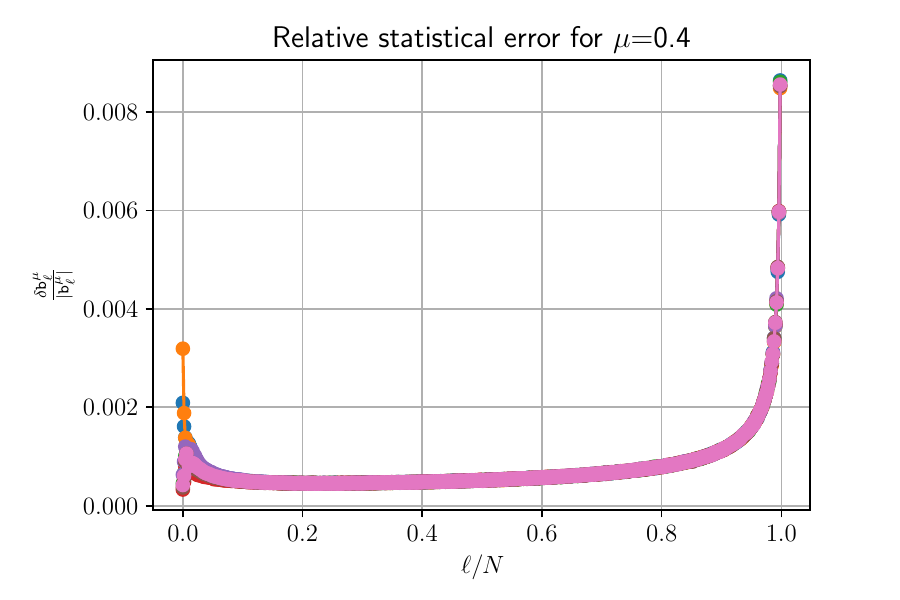}
         \caption{$\mu = 0.4$}
     \end{subfigure}   
        \hfill
    \begin{subfigure}[b]{0.32\textwidth}
        \centering         \includegraphics[width=\textwidth]{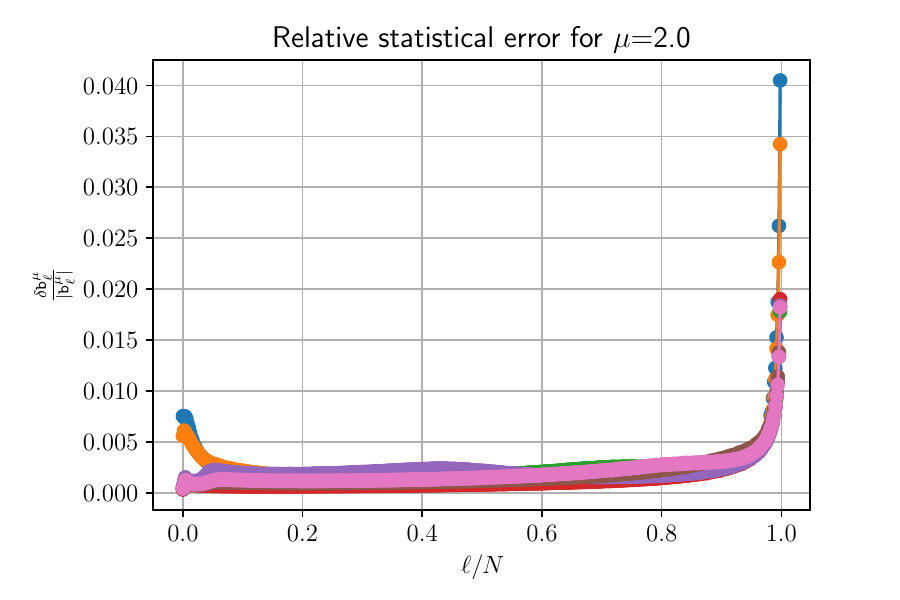}
          \caption{$\mu = 2$}
     \end{subfigure}   
        \hfill
    \begin{subfigure}[b]{0.32\textwidth}
        \centering
         \includegraphics[width=\textwidth]{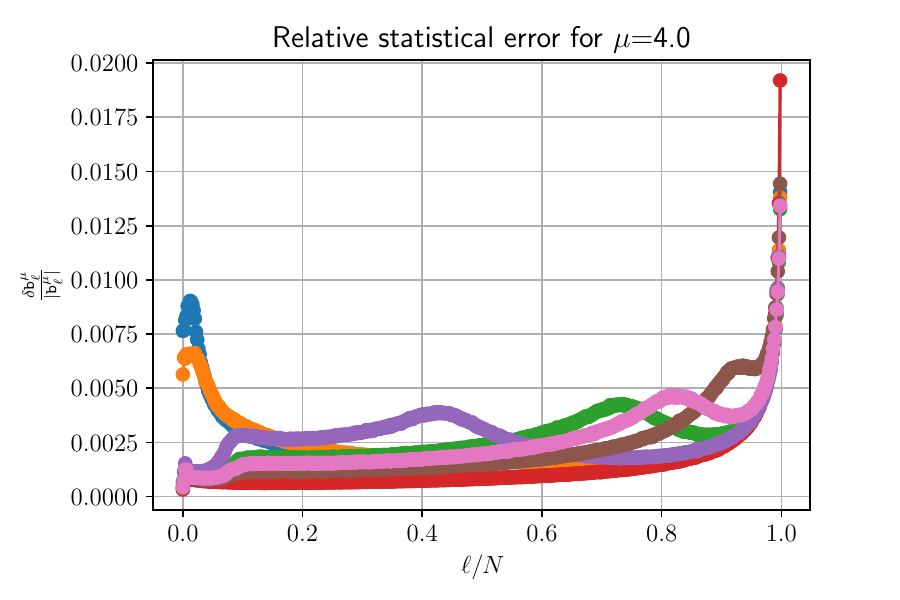}
          \caption{$\mu =4 $}
     \end{subfigure}  
    \caption{The relative error of LCD $\mathtt{b}_\ell^\mu$ for all initial states with $\mu = \{0.4, 2, 4\}$. The matrix size is $N=500$ and the ensemble size is $M=9600$.
} \label{fig:relative_std_vs_k_N_500}
\end{figure}

To illustrate, we have plotted the relative statistical error for the $\mathtt{b}^\mu_\ell$ distribution of the initial states in Fig. \ref{fig:relative_std_vs_k_N_500}. We see that the error at the tail region gets enhanced but it is still small. Due to this enhancement at the tail region where $\ell \sim N$, the error of the higher moments will be bigger. As for the $\mu$ dependence, the statistical error grows with $\mu$, reaches the maximum, and then decreases.

\begin{figure}[!h]
    \begin{subfigure}[b]{0.32\textwidth}
        \centering
         \includegraphics[width=\textwidth]{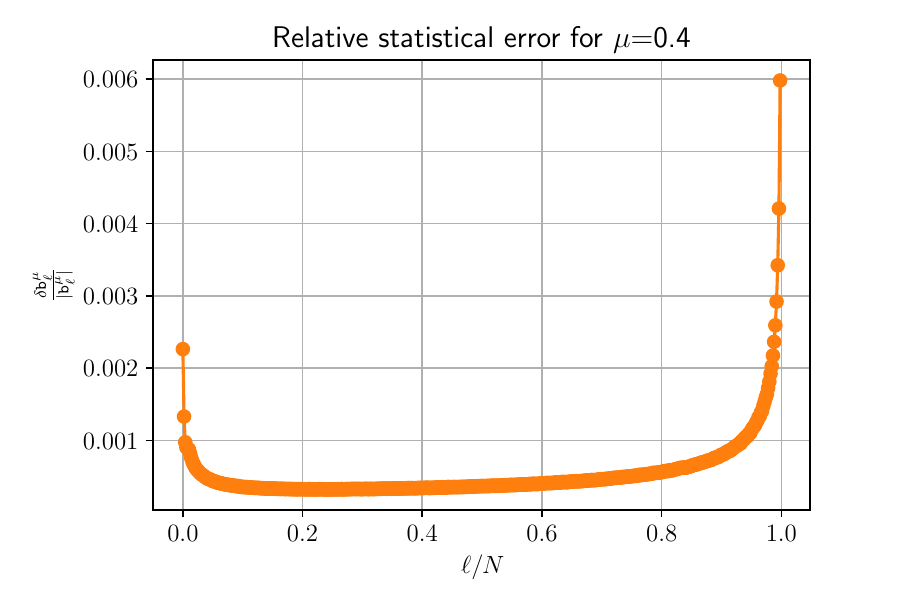}
         \caption{$\mu = 0.4$}
     \end{subfigure}   
        \hfill
    \begin{subfigure}[b]{0.32\textwidth}
        \centering         \includegraphics[width=\textwidth]{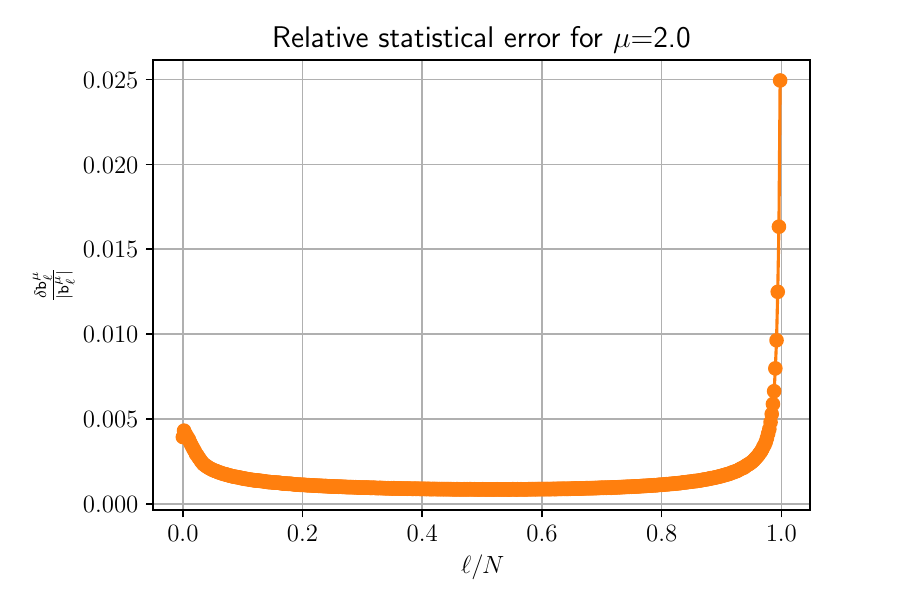}
          \caption{$\mu = 2$}
     \end{subfigure}   
        \hfill
    \begin{subfigure}[b]{0.32\textwidth}
        \centering
         \includegraphics[width=\textwidth]{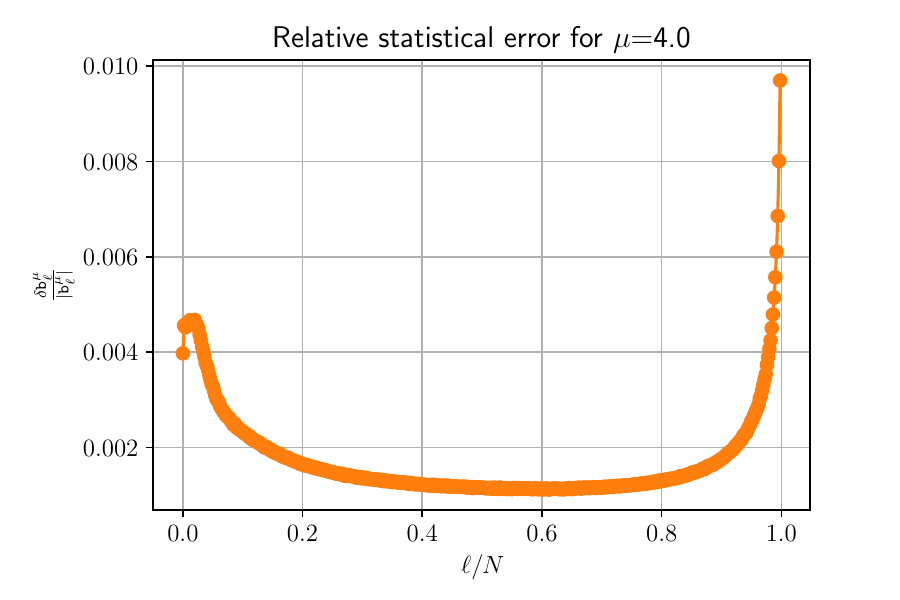}
          \caption{$\mu =4 $}
     \end{subfigure}  
    \caption{The relative error of LCD $\mathtt{b}_\ell^\mu$ for initial state $\textcolor{mOrange}{|\psi_2\rangle}$ with $\mu = \{0.4, 2, 4\}$. The matrix size is $N=500$ and the ensemble size is $M=19200$.
} \label{fig:relative_std_vs_k_psi2}
\end{figure}

We have also plotted the relative statistical error for the $\mathtt{b}^\mu_\ell$ distribution of the initial state $|\psi_2\rangle$ in Fig. \ref{fig:relative_std_vs_k_psi2}. The error is smaller than the previous plot due to the larger ensemble size, but the qualitative behavior remains the same.

Based on the LCD errors, we can now plot the LCD moments and the LCD entropy with error bars, as shown in Fig. \ref{fig:bn_states_moments}, \ref{fig:bn_states_entropy}, \ref{fig:bn_states_moments_psi2} and \ref{fig:bn_states_entropy_psi2}, respectively. Errors of the subsequent fittings depicted in Figs. \ref{fig:bn_prob_slope_vs_moment} and \ref{fig:bn_prob_alpha_vs_moment_psi2} can then be calculated by the data errors. We use the $\mathtt{SciPy}$ package, $\mathtt{scipy.optimize.curve\_fit}$ function to perform the fit, and the error estimates for the parameters are built-in.

Now we address issues that are not directly resolved by error analysis.
\paragraph{Identifying $\mu_c$ Range.}

\begin{figure}[!h]
    \begin{subfigure}[b]{0.48\textwidth}
        \centering
         \includegraphics[width=\textwidth]{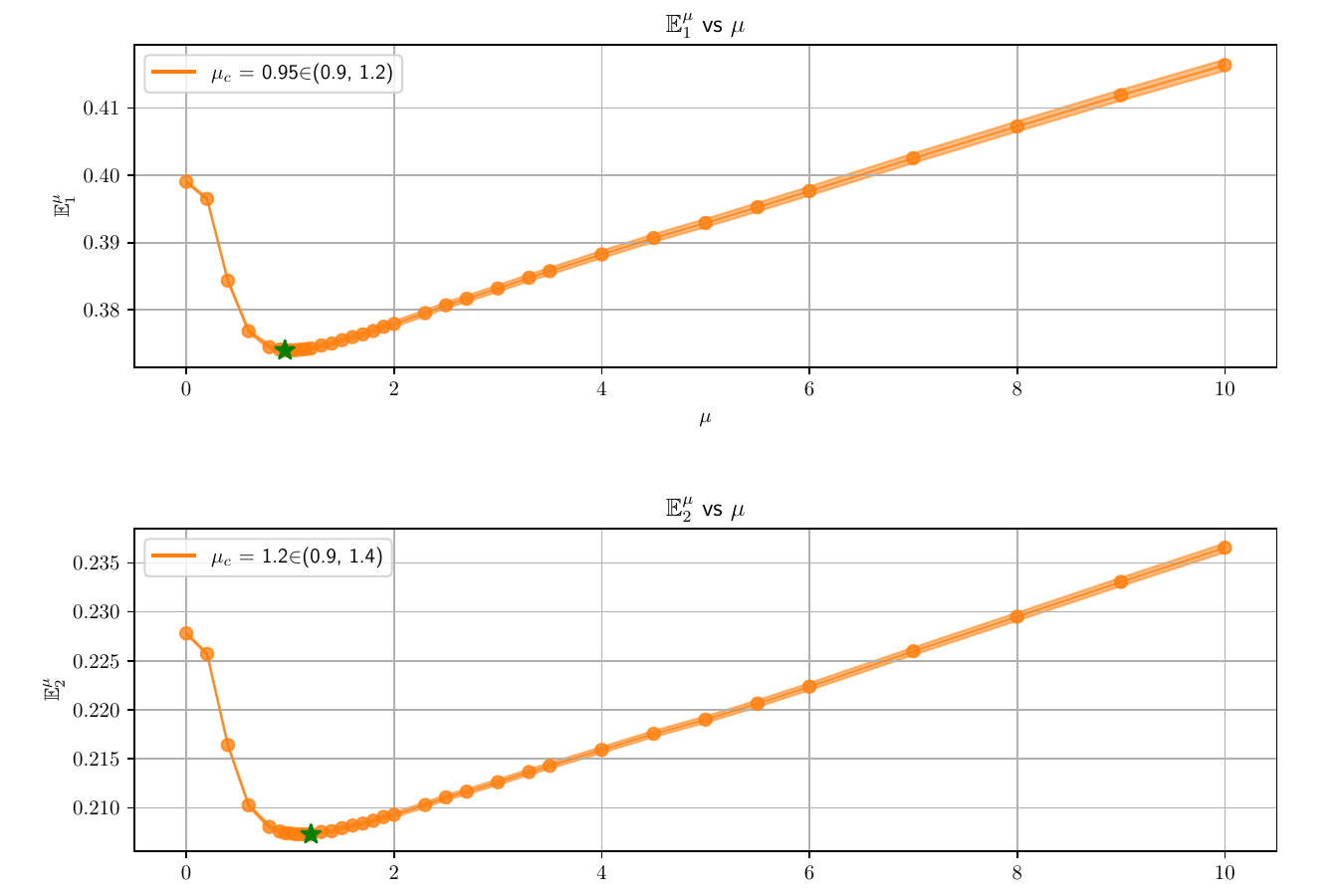}
         \caption{$\gamma = 1$}
     \end{subfigure}  
     \hfill
    \begin{subfigure}[b]{0.48\textwidth}
        \centering         \includegraphics[width=\textwidth]{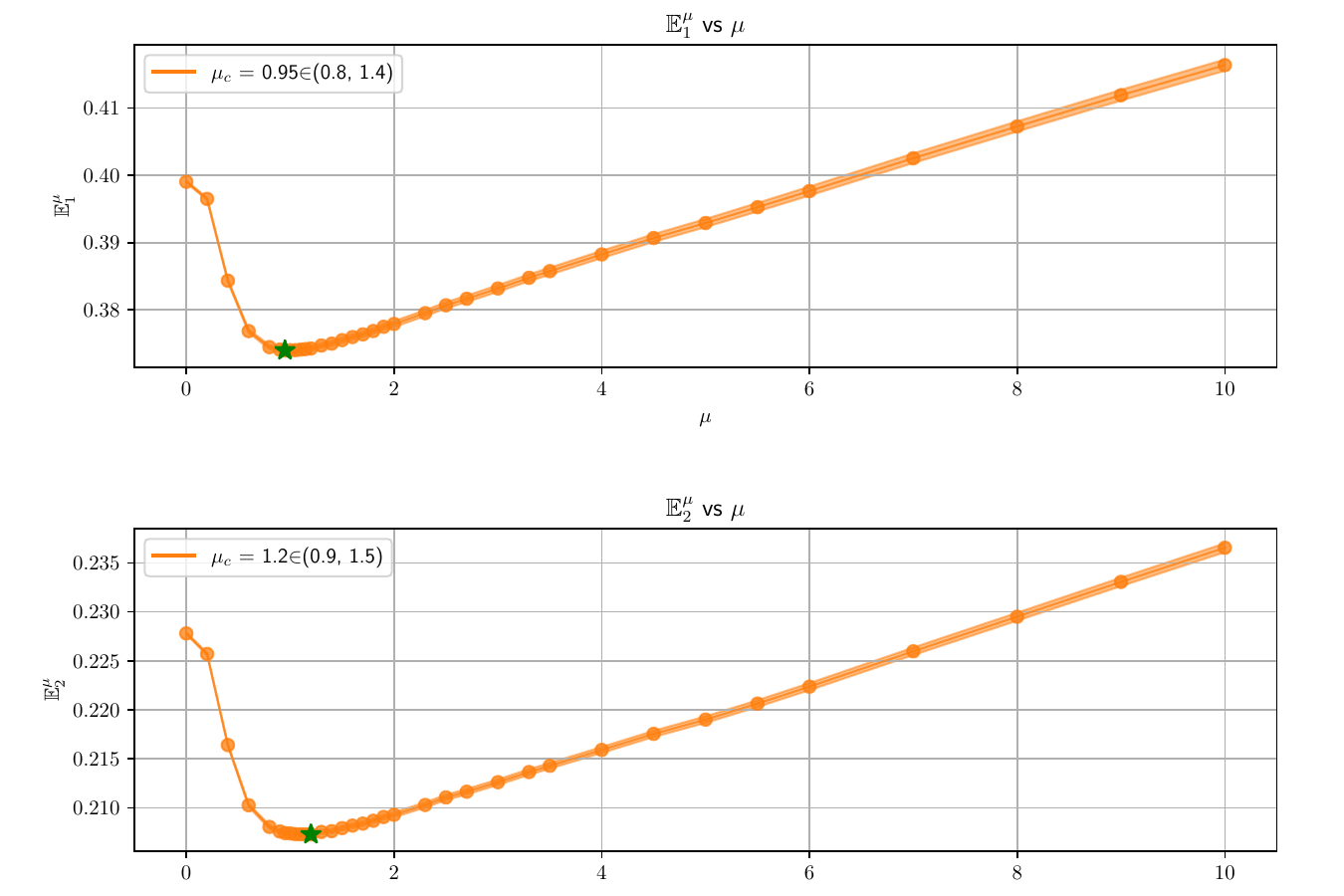}
          \caption{$\gamma = 2$}
     \end{subfigure}   
    \caption{An alternative method of choosing the $\mu_c$ range: we take all points near $\mu_c$ such that $\mathbb{E}_k^{\mu} - \mathbb{E}_k^{\mu_c} < \gamma \sqrt{(\delta \mathbb{E}_k^{\mu})^2 + (\delta \mathbb{E}_k^{\mu_c})^2}$.
} \label{fig:mucRangeOtherMethods}
\end{figure}

We define $\mu_c$ as the minimum point of the curve $\mathbb{E}_k^{\mu}$. However, due to the error distribution, the minimum point is actually a random variable. We bound the distribution of the minimum point by taking the minimum and the maximum value from the set of all the minimum $\mu$ candidates, as can be shown in Fig. \ref{fig:bn_states_moments_psi2}.The set of all the possible $\mu$ values is defined as all the $\mu$ such that $\mathbb{E}_k^{\mu} - \delta \mathbb{E}_k^{\mu} < \mathbb{E}_k^{\mu_c} + \delta \mathbb{E}_k^{\mu_c}$. 
The criterion of selecting the $\mu_c$ region is not unique. For instance, we can select all $\mu$ points such that $\mathbb{E}_k^{\mu} - \mathbb{E}_k^{\mu_c} < \gamma \sqrt{(\delta \mathbb{E}_k^{\mu})^2 + (\delta \mathbb{E}_k^{\mu_c})^2}$, with $\gamma = 1$ or $2$. The qualitative behavior of the moments remains the same, see Fig. \ref{fig:mucRangeOtherMethods} for an illustration. 

\paragraph{Large $\mu$ Fit.}

The algorithm for fitting Fig. \ref{fig:bn_prob_slope_vs_moment} is the following: we start to fit the straight line by using the last three points. After the fitting process, we calculate the statistical error associated with the fitted slope. Next, we add an additional point to the fit and proceed with the same fitting process, continuing until the statistical error of the fitted slope begins to rise.

\paragraph{$\alpha$ Fit.}

The $\alpha$ fit in Fig. \ref{fig:bn_prob_alpha_vs_moment_psi2} has an ambiguity in identifying the precise location of $\mu_c$. We have taken the data minimum to be $\mu_c \simeq 1.2$ in the main text, but other values of $\mu_c$ in the obtained error range also give smooth plots.

\section{R\'{e}nyi Entropy of the LCD Distribution} \label{apd:RenyiEntropy}

In Fig. \ref{fig:EntropyRenyi}, we have plotted the LCD R\'{e}nyi entropy for $\alpha = 1, 4$ and $6$. The qualitative behaviors are identical to Fig. \ref{fig:bn_states_entropy}. However, for small $\alpha$ we observe some crossover of the $\textcolor{mPurple}{|\psi_3\rangle}$. 
Moreover, the crossover of the $\textcolor{mPurple}{|\psi_3\rangle}$ state disappears for $\alpha >6$, suggesting that this crossover is non-physical. 
\begin{figure}[!ht]
    \begin{subfigure}[b]{0.32\textwidth}
        \centering
         \includegraphics[width=\textwidth]{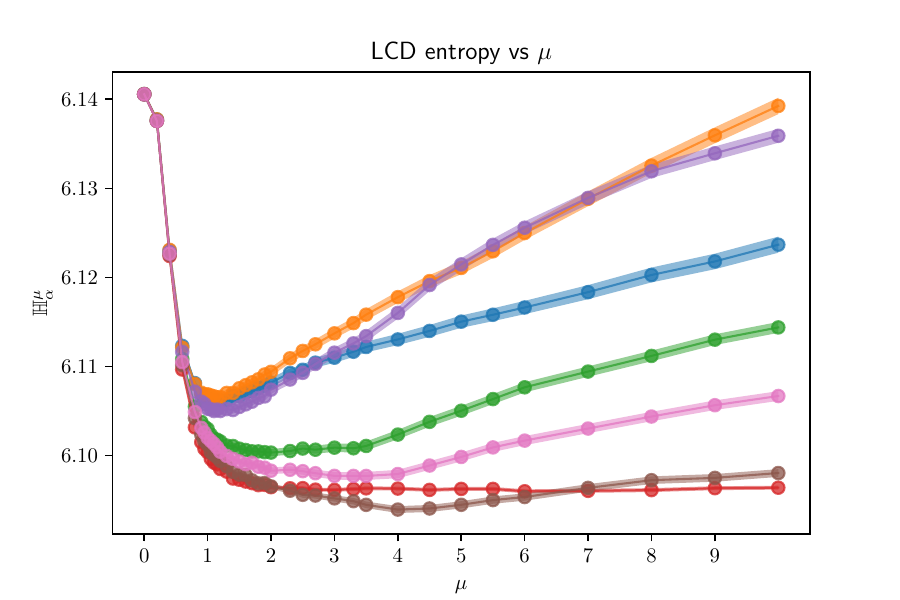}
         \caption{$\alpha = 1$}
     \end{subfigure}   
        \hfill
    \begin{subfigure}[b]{0.32\textwidth}
        \centering         \includegraphics[width=\textwidth]{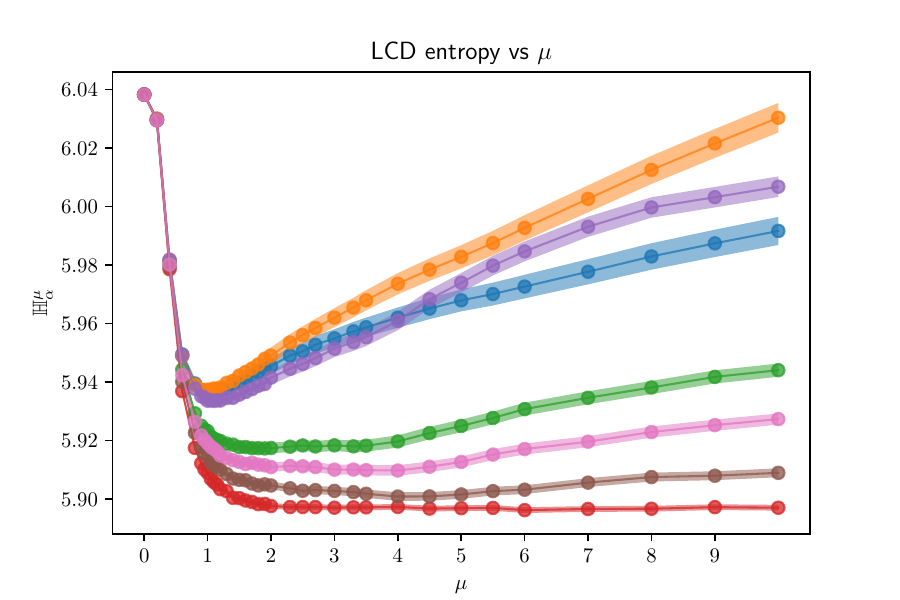}
          \caption{$\alpha = 4$}
     \end{subfigure}   
        \hfill
    \begin{subfigure}[b]{0.32\textwidth}
        \centering
         \includegraphics[width=\textwidth]{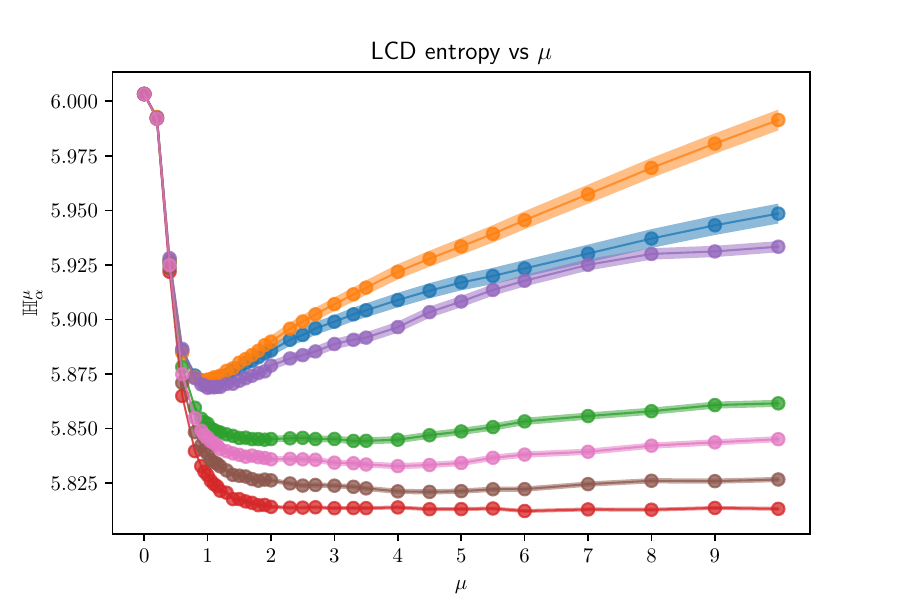}
          \caption{$\alpha = 6$}
     \end{subfigure}  
\caption{LCD R\'{e}nyi entropy with $\alpha = \{1, 4, 6\}$. The matrix size is $N=500$ and the ensemble size is $M=9600$.}  
\label{fig:EntropyRenyi}
\end{figure}

To determine the transition region, we present the results of the LCD R\'{e}nyi entropy for state $\textcolor{mOrange}{|\psi_2\rangle}$. We observe that the statistical error decreases in the large $\mu$ region as $\alpha$ increases. 

When increasing $\alpha$, the value of the minimum point $\mu_c$ decreases. This behavior aligns with LCD predictions, where higher $\alpha$ values receives dominant contributions from the large $\mathtt{b}_\ell$ region, corresponding to the small $\ell/N$ region in our LCD case.

\begin{figure}[!h]
    \begin{subfigure}[b]{0.32\textwidth}
        \centering
        \includegraphics[width=\textwidth]{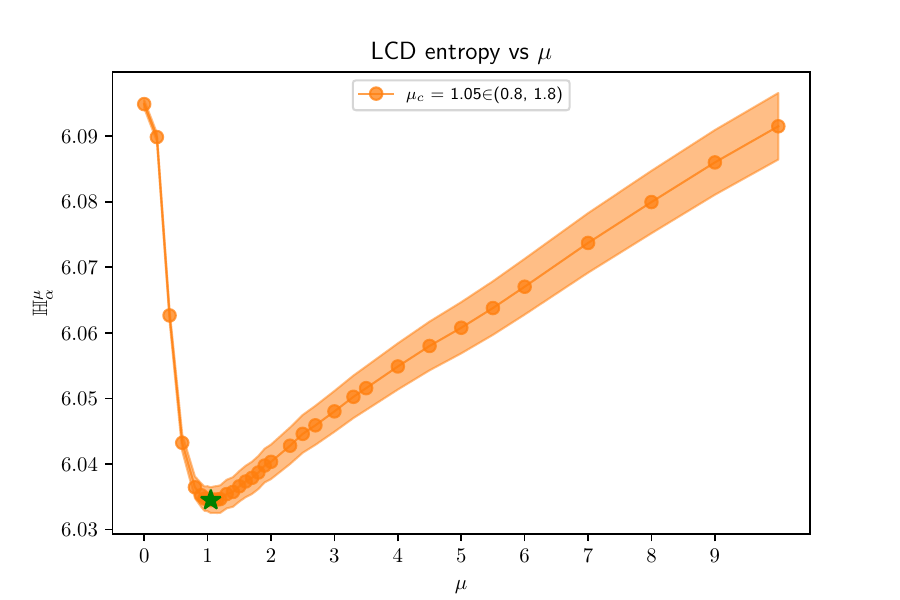}
         \caption{$\alpha = 2$}
     \end{subfigure}   
        \hfill
    \begin{subfigure}[b]{0.32\textwidth}
        \centering      
    \includegraphics[width=\textwidth]{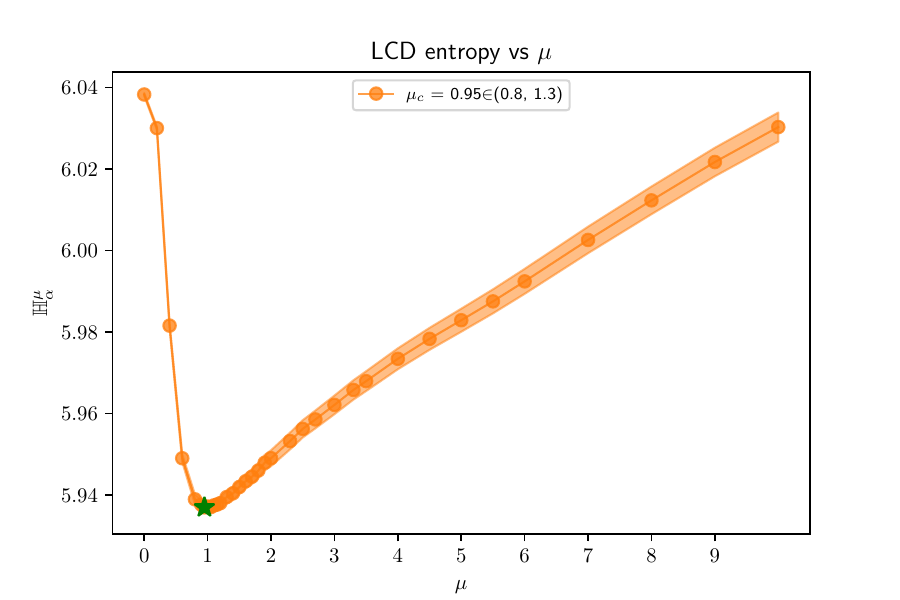}
          \caption{$\alpha = 4$}
     \end{subfigure}   
        \hfill
    \begin{subfigure}[b]{0.32\textwidth}
        \centering
         \includegraphics[width=\textwidth]{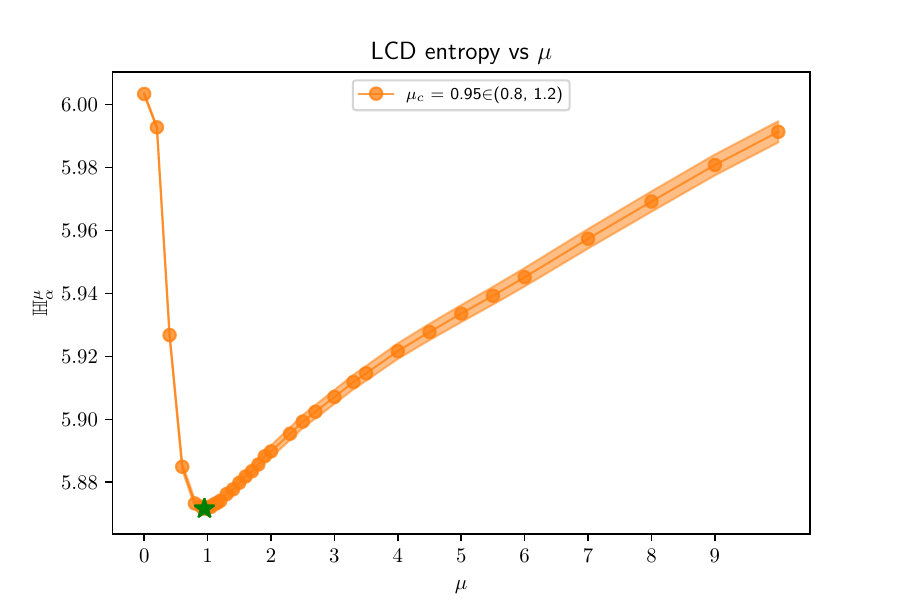}
          \caption{$\alpha = 6$}
     \end{subfigure}  
\caption{LCD R\'{e}nyi entropy with $\alpha = \{2, 4, 6\}$. The matrix size is $N=500$ and the ensemble size is $M=19200$.}  
\label{fig:EntropyRenyi}
\end{figure}

\section{An Alternative Fit of the Moments}
\label{sec:altenativeFit}

\begin{figure}[!h]
    \centering \includegraphics[width = 0.5\textwidth]{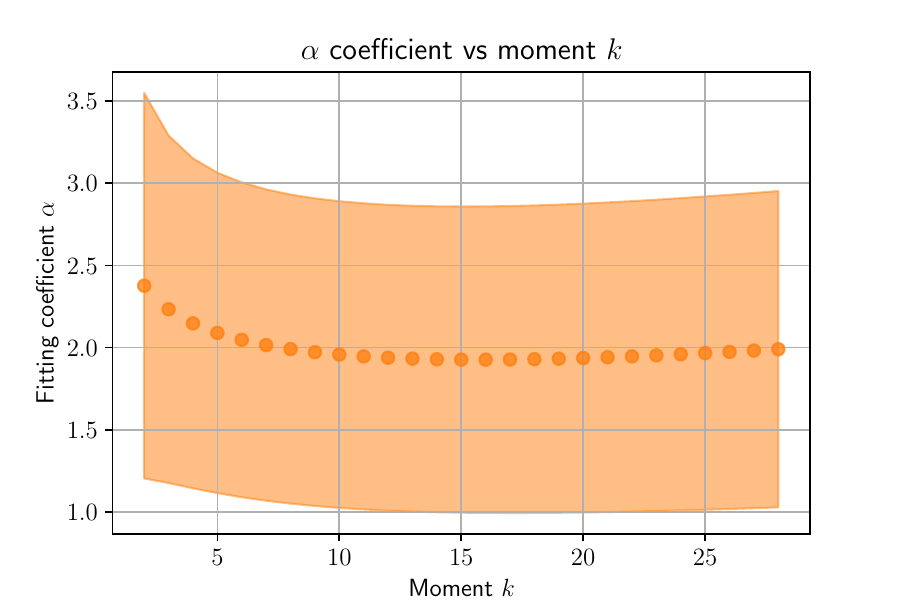}
    \caption{The power-law fit near the critical point for initial state $\textcolor{mOrange}{|\psi_2\rangle}$.}
    \label{fig:AlphaTaylor_Psi2}
\end{figure}

In the transition region discussed in section \ref{sec:transition}, we have chosen the quadratic Taylor expansion \eqref{scaling} to fit the data. The reason for doing that is because such a fit has reasonable errors. To compare, we consider an alternative power-law fit near the critical point, described by:
\begin{equation}\label{scaling2}
\mathbb{E}_k^{\mu} \sim (\mu-\mu_c)^{\alpha_k
} \ ,
\end{equation}
as depicted in Fig. \ref{fig:AlphaTaylor_Psi2}. We see that the error is of the same order as for the data points, thus the proposed functional form \eqref{scaling2} is not suitable for describing the vicinity of $\mu_c$.

\section{Unnormalized Lanczos Coefficients Dependence on $\mu$} \label{apd:Unnormalized}

\begin{figure}[!h]
    \begin{subfigure}[b]{0.45\textwidth}
        \centering
         \includegraphics[width=\textwidth]{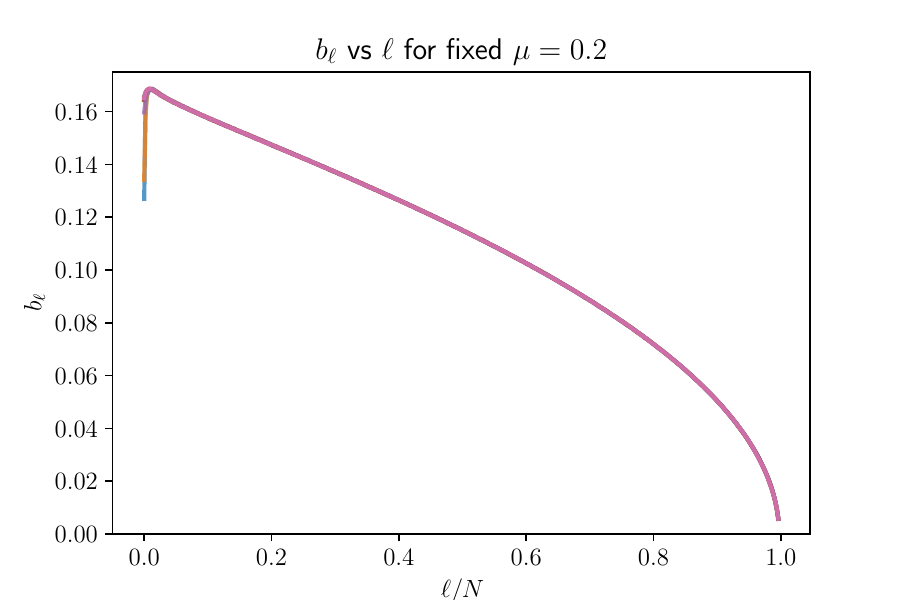}
         \caption{For fixed $\mu=0.2$ before the transition point $\mu_c \simeq 1.2$, the unnormalized $b_\ell$ as a function of different initial states. There is no dependence on the initial state, as expected.}
         \label{fig:bl_vs_mu_beforeTran}
     \end{subfigure}   
        \hfill
    \begin{subfigure}[b]{0.45\textwidth}
         \centering
         \includegraphics[width=\textwidth]{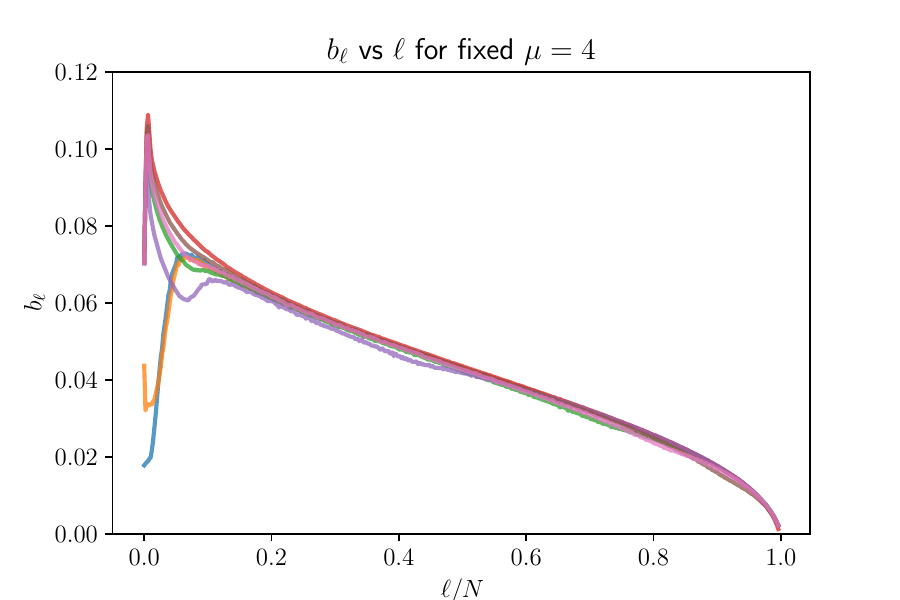}
         \caption{For fixed $\mu=4$ after the transition point $\mu_c \simeq 1.2$, the unnormalized $b_\ell$ as a function of different initial states. The $b_\ell$ data shows dependence on the initial state, mostly for the small $\ell/N$ range.}
         \label{fig:bl_vs_mu_afterTran}
     \end{subfigure}   
\caption{$b_\ell$ dependence on $\ell$ for fixed $\mu$ before and after the transition.}
    \label{fig:bl_vs_ell}
\end{figure}

This section presents the raw data for the Lanczos coefficients of $b_\ell$.
Recall that the data points for $b_\ell$ depend on the initial state and the parameter $\mu$. To illustrate this dependency more effectively, we will demonstrate it through two separate figures.
In Fig. \ref{fig:bl_vs_ell}, we depict the $b_\ell$ values for all different initial states, keeping $\mu$ fixed. We find that for $\mu < \mu_c \simeq 1.2$ the $b_\ell$ values show no dependence on the initial states. Conversely, for $\mu >\mu_c \simeq 1.2$, a distinct dependence on the initial state becomes evident, especially for the $\ell/N$ small region.

\begin{figure}[!h]
    \centering
    \includegraphics[width=0.5\linewidth]{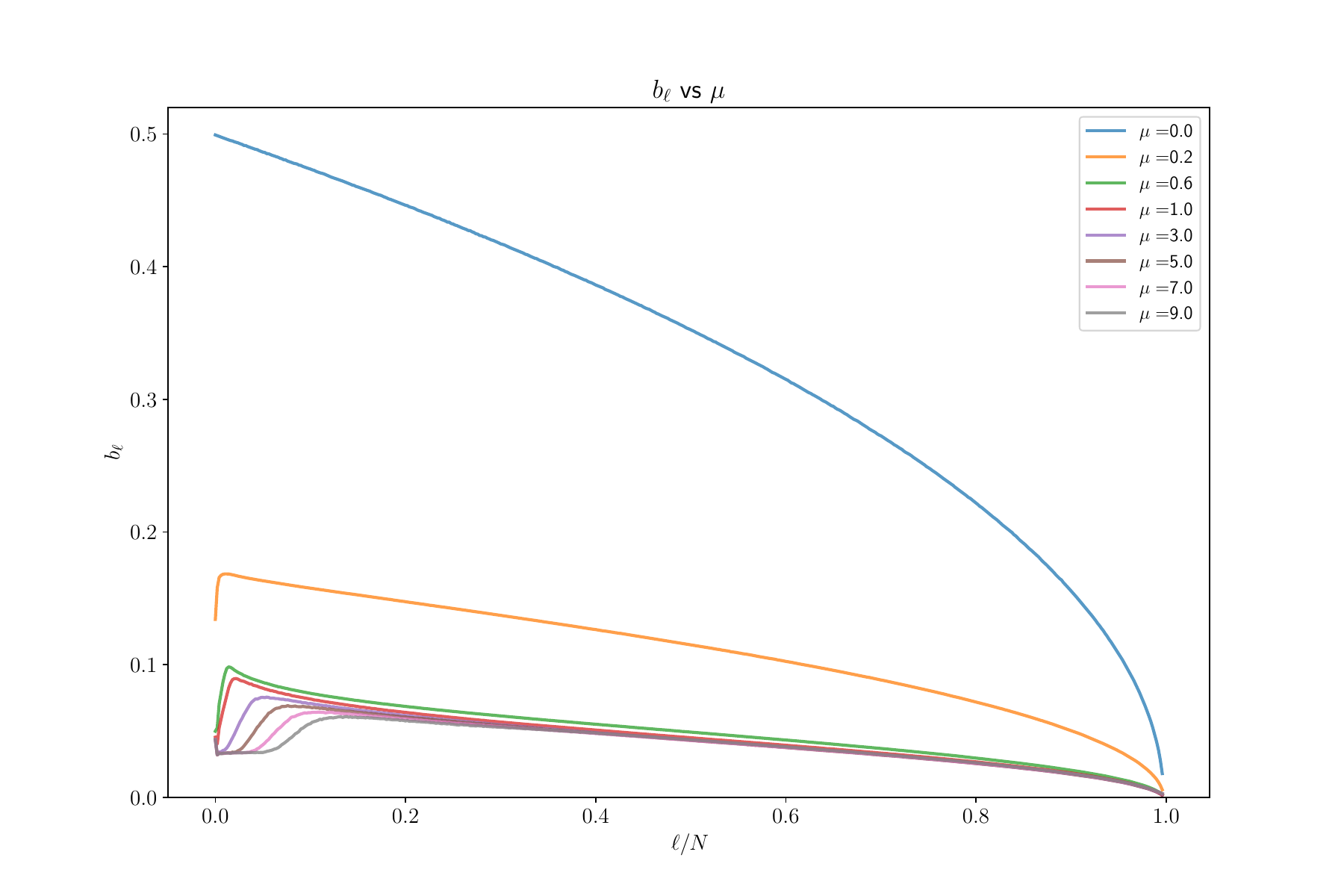}
    \caption{The unnormalized $\langle b_\ell \rangle$ as a function of $\mu$ for initial state $|\psi_2\rangle$. Here the colors represent different $\mu$ rather than initial states.}
    \label{fig:bl_vs_mu_psi2}
\end{figure}

In Fig. \ref{fig:bl_vs_mu_psi2}, we have plotted the dependence of $b_\ell$ on $\mu$ for a specific initial state $|\psi_2\rangle$. Given the similarity in the $b_\ell$-$\mu$ relationship among different initial states, we present just this single case for clarity. One finds that the magnitude of $b_\ell$ decreases as one increases $\mu$. However, there is no clear implication from the plot near the transition point $\mu_c \simeq 1.2$. 

\begin{figure}[hbt!]
 
    \includegraphics[width=0.5\textwidth]{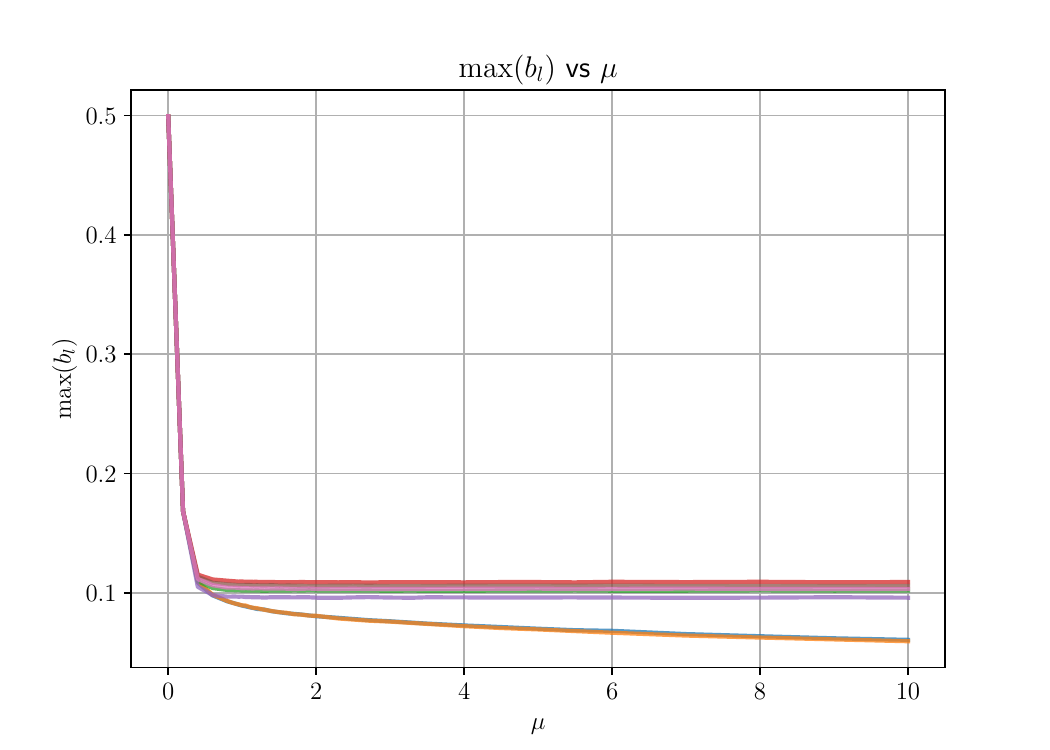}
    \caption{The maximal values of $b_\ell$ (note there are not normalized) as functions of the correlation strength $\mu$ are represented by the \textcolor{mBlue}{blue}, \textcolor{mOrange}{orange}, \textcolor{mPurple}{purple}, \textcolor{mGreen}{green}, 
    \textcolor{mPink}{pink},
    \textcolor{mBrown}{brown} and \textcolor{mRed}{red} data points correspond to the initial states $\textcolor{mBlue}{|\psi_1\rangle}, \cdots \textcolor{mRed}{|\psi_7\rangle}$ defined in \eqref{eqn-IntialState} respectively.}
    \label{fig:bl_max}
\end{figure}

The only evident observable that makes a difference for the transition is the maximal values of $b_\ell$, shown in Fig. \ref{fig:bl_max}. We observe, qualitatively, that the dependence of $\max(b_\ell)$ on $\mu$ is linear. However, there is a noticeable change in slope near the transition point, $\mu_c \approx 1$.
The analysis above shows the importance of normalization in defining the LCD distribution. It is only after applying the correct normalization that $b_\ell$ becomes a well-defined probability distribution, enabling us to obtain quantitative information about the phase transition.

\section{Size Dependence of the LCD Moments and R\'{e}nyi Entropy }
\label{apd:Ndep}

This section presents the LCD moments and the R\'{e}nyi entropy at $\alpha =10$ for various matrix sizes. The purpose is to demonstrate that at the matrix size $N=500$, as used in the main text, the physical system has reached the scaling limit.

\begin{figure}[hbt!]
    \centering
    \includegraphics[width=0.6\linewidth]{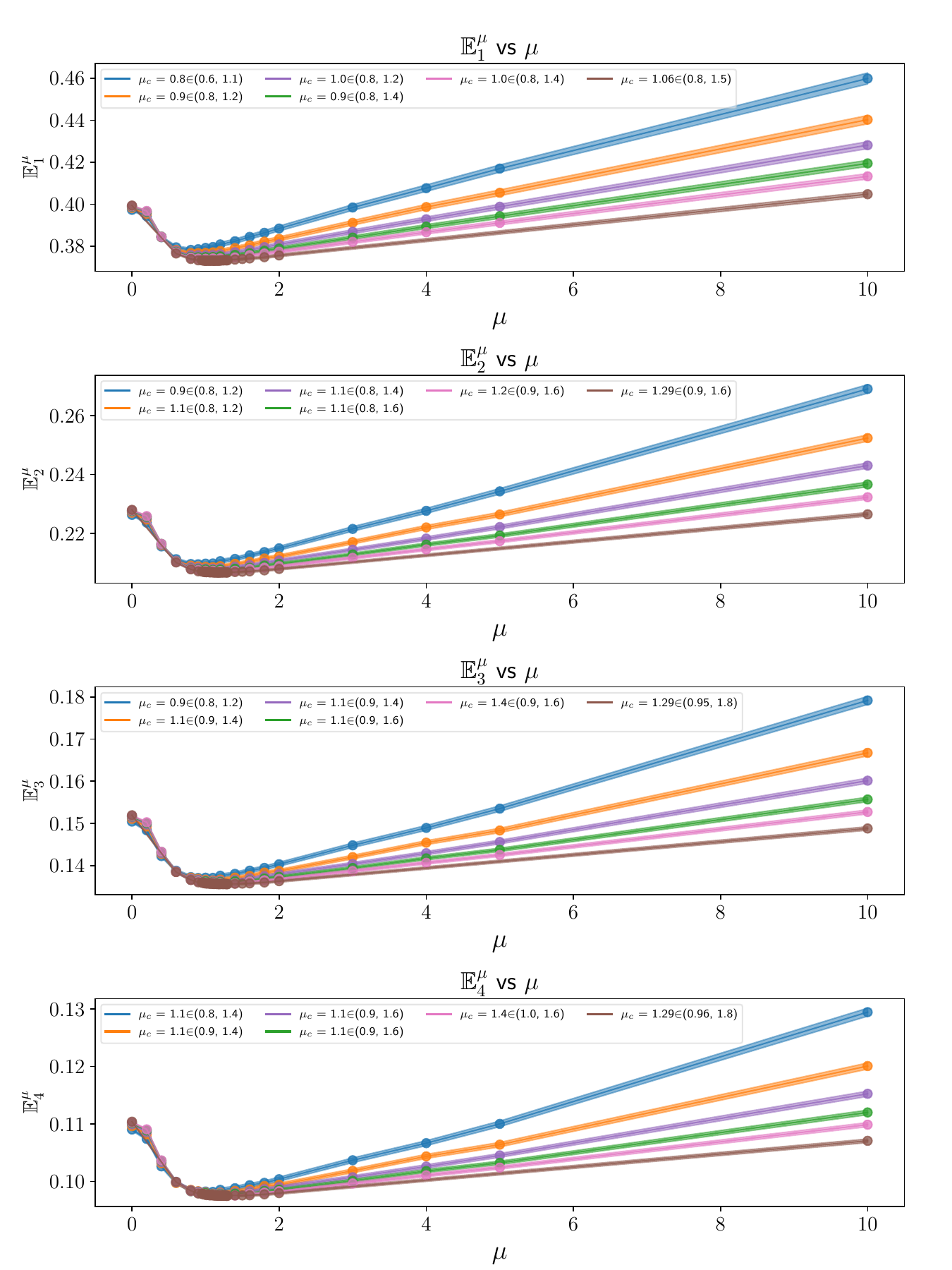}
    \caption{The first four LCD moments as functions of $N \in \{\textcolor{mBlue}{200}, \textcolor{mOrange}{300}, \textcolor{mPurple}{400}, \textcolor{mGreen}{500}, 
    \textcolor{mPink}{600},
    \textcolor{mBrown}{800} \}$ for initial state ${|\psi_2\rangle}$ with error bars.
    The ensemble size is $M=6000$. }
    \label{fig:prob_N_dep}
\end{figure}

In Figure \ref{fig:prob_N_dep}, we have plotted the first four LCD moments for the initial state  $|\psi_2\rangle$ for various matrix sizes, ranging from  $N=200$ to $N=800$. Unlike in other sections, here the colors represent different matrix sizes rather than different initial states.

From the figure, we can observe that the statistical error bars are very small; the primary differences arise from the matrix size. The largest discrepancies occur in the large  $\mu$  region. As shown in Table \ref{tab:NDepE}, the relative difference in the moment values decreases to less than two percent once the matrix size reaches $N=500$.

\begin{figure}[hbt!]
    \centering
    \includegraphics[width=0.5\linewidth]{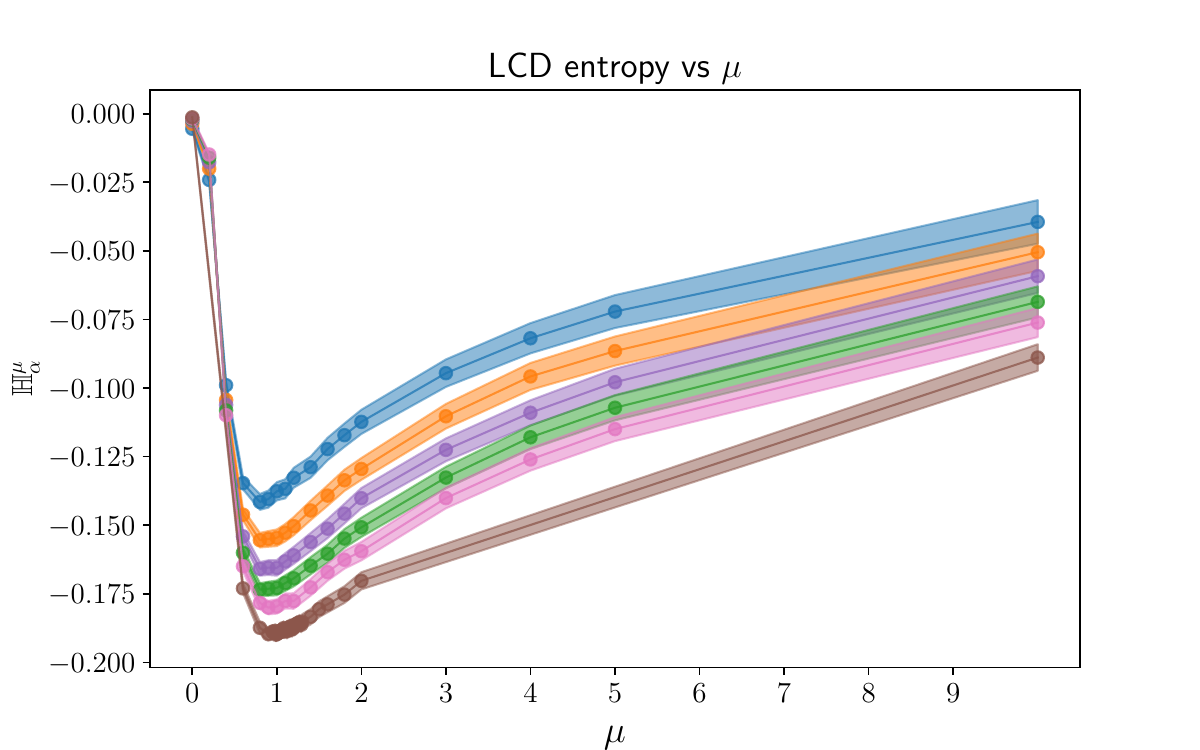}
    \caption{The R\'{e}nyi Entropy at $\alpha =10$ as functions of $N \in \{\textcolor{mBlue}{200}, \textcolor{mOrange}{300}, \textcolor{mPurple}{400}, \textcolor{mGreen}{500}, 
    \textcolor{mPink}{600},
    \textcolor{mBrown}{800} \}$ for initial state ${|\psi_2\rangle}$ with error bars. For comparison, we have subtracted the GOE value from each of them.
    The ensemble size is $M=6000$. }
    \label{fig:Entropy_N_dep}
\end{figure}

Importantly, across the different ranges of  $N$  used, the qualitative behavior of the LCD moments remains the same: in the ergodic regime, all moments are independent of  $\mu$ , while for  $\mu$  greater than one, the dependence on the initial state becomes noticeable. Combined with the error analysis above, we believe that the system size we use has reached the scaling limit. Similar conclusions can be drawn from the R\'{e}nyi entropy, see Figure \ref{fig:Entropy_N_dep}.

\begin{table}[h!]
\centering
\begin{tabular}{|c|c|c|c|c|c|c|c|c|}
\hline
$N$ & $\mathbb{E}^{\mu=10}_1$ & $|\delta \mathbb{E}^{\mu=10}_1|/\mathbb{E}^{\mu=10}_1$ & $\mathbb{E}^{\mu=10}_2$ & $|\delta \mathbb{E}^{\mu=10}_2|/\mathbb{E}^{\mu=10}_2$ & $\mathbb{E}^{\mu=10}_2$ & $|\delta \mathbb{E}^{\mu=10}_3|/\mathbb{E}^{\mu=10}_3$ & $\mathbb{E}^{\mu=10}_3$ & $|\delta \mathbb{E}^{\mu=10}_4|/\mathbb{E}^{\mu=10}_4$ \\ \hline
200 & 0.4599 & 0.0427 & 0.2690 & 0.0618 & 0.1792 & 0.0694 & 0.1295 & 0.0724 \\ \hline
300 & 0.4402 & 0.0276 & 0.2524 & 0.0369 & 0.1667 & 0.0396 & 0.1201 & 0.0402 \\ \hline
400 & 0.4281 & 0.0203 & 0.2430 & 0.0263 & 0.1601 & 0.0279 & 0.1153 & 0.0282 \\ \hline
500 & 0.4194 & 0.0147 & 0.2367 & 0.0183 & 0.1557 & 0.0190 & 0.1120 & 0.0190 \\ \hline
600 & 0.4133 & 0.0205 & 0.2323 & 0.0250 & 0.1527 & 0.0257 & 0.1099 & 0.0256 \\ \hline
800 & 0.4047 & - & 0.2265 & - & 0.1488 & - & 0.1071 & - \\ \hline
\end{tabular}
\caption{
The values of  $\mathbb{E}_i^\mu$  for  $\mu = 10$  and  $i = 1,2,3,4$  are presented below. Since the statistical error is significantly smaller than the moment values, we only display the central values of  $\mathbb{E}_i^\mu$  in the table.
} \label{tab:NDepE}
\end{table}
 
\end{document}